\documentclass[11pt,a4paper]{article}
\pdfoutput=1
\usepackage{jcappub}
\usepackage{amsmath}
\usepackage{graphicx}
\usepackage{color}
\usepackage{natbib}
\usepackage{aas_macros} 
\usepackage{bm}

\makeatletter
\g@addto@macro\bfseries{\boldmath}
\makeatother

\newcommand{\di}{\text{d}}
\newcommand{\diff}[2]{\frac{\di #1}{\di #2}}
\newcommand{\diffpar}[2]{\frac{\partial #1}{\partial #2}}
\newcommand{\vek}[1] {{\bf #1} }
\renewcommand{\vec}{\vek}
\newcommand{\Hc}{\mathcal{H}}

\newcommand{\Fourier}[2]{\mathcal{F}_{\bf #1}\left[ #2 \right]}
\newcommand{\FourierMap}{\xrightarrow{\mathcal{F}_\vec{k}}}
\newcommand{\Kernel}[1]{\mathcal{C}_\vec{k}\left\{ #1 \right\}}
\newcommand{\KernelPhi}[1]{\mathcal{C}_\vec{k}\left\{\phi_0\left( \vek{k}_1 \right) \phi_0\left( \vek{k}_2 \right) #1 \right\}}
\newcommand{\KernelDelta}[2]{\mathcal{C}_\vec{k}\left\{ \frac{9 \Hc^4_0 \Omega_m^2}{4#1}  \frac{\delta ( \vek{k}_1 )}{1+3 f \frac{\Hc^2}{k_1^2}} \frac{\delta ( \vek{k}_2 )}{1+3f \frac{\Hc^2}{k_2^2}} #2 \right\}}
\newcommand{\KernelDeltaBis}[2]{\mathcal{C}_\vec{k}\left\{ \frac{9 \Hc^4_0 \Omega_m^2}{4#1}  \frac{\delta ( \vek{k}_1 )}{1+3 f \frac{\Hc^2}{k_1^2}} \frac{\delta ( \vek{k}_2 )}{1+3f\frac{\Hc^2}{k_2^2}}  #2 \right.}
\newcommand{\coskk}{\hat{\vek{k}}_1 \cdot \hat{\vek{k}}_2}
\newcommand{\HcO}{H_0^2 \Omega_{m}}
\newcommand{\Hcdot}{\dot{\mathcal{H}}}
\newcommand{\SONG}{\textsc{song}}
\newcommand{\CLASS}{\textsc{class}}
\newcommand{\etilde}{\tilde{\eta}}
\newcommand{\xtilde}{\tilde{x}}

\newcommand{\dlogpar}[2]{\frac{\partial \log #1}{\partial \log #2}}
\newcommand{\hypF}[4]{\,{_2}F_1 \left( #1, #2, #3, #4 \right)}
\newcommand{\fref}[1]{figure~\ref{#1}}
\newcommand{\velo}{\boldsymbol{\mathcal{U}}}
\newcommand{\kinavg}[1]{\left\langle #1 \right\rangle_{\vec{u}}}
\newcommand{\velpot}{\Upsilon}
\newcommand{\deltil}{\tilde{\delta}}
\newcommand{\delone}{\delta^{(1)}}

\usepackage[normalem]{ulem}

\begin{document}
 
\title{The Intrinsic Matter Bispectrum in $\Lambda$CDM}

\author[a]{Thomas Tram,}
\emailAdd{thomas.tram@port.ac.uk}
\author[b]{Christian Fidler,}
\author[a]{Robert Crittenden,}
\author[a]{Kazuya Koyama,}
\author[a,c]{Guido W. Pettinari}
\author[a]{and David Wands}

\affiliation[a]{Institute of Cosmology and Gravitation, University of Portsmouth, Portsmouth PO1 3FX, United Kingdom}
\affiliation[b]{Catholic University of Louvain - Center for Cosmology, Particle Physics and Phenomenology (CP3)
2, Chemin du Cyclotron, B-1348 Louvain-la-Neuve, Belgium}
\affiliation[c]{Department of Physics \& Astronomy,
University of Sussex, Brighton BN1 9QH, UK}

\date{\today}

\abstract{We present a fully relativistic calculation of the matter bispectrum at second order in cosmological perturbation theory assuming a Gaussian primordial curvature perturbation.  For the first time we perform a full numerical integration of the bispectrum for both baryons and cold dark matter using the second-order Einstein-Boltzmann code, \SONG{}. We review previous analytical results and provide an improved analytic approximation for the second-order kernel in Poisson gauge which incorporates Newtonian nonlinear evolution, relativistic initial conditions, the effect of radiation at early times and the cosmological constant at late times. Our improved kernel provides a percent level fit to the full numerical result at late times for most configurations, including both equilateral shapes and the squeezed limit.  We show that baryon acoustic oscillations leave an imprint in the matter bispectrum, making a significant impact on squeezed shapes. }

\maketitle

\section{Introduction}

The Large Scale Structure (LSS) of the Universe is one of the most promising cosmological probes. Cosmic Microwave Background (CMB) observations provide the most precise current measurements of primordial perturbations, but they probe primarily the two-dimensional last-scattering sphere and the constraints placed upon theoretical models are thus reaching the cosmic-variance limit \cite{Ade:2015xua}. In contrast, LSS experiments can probe many independent observables at many different redshifts. The next generation of experiments \cite{laureijs2011euclid} will surpass the sensitivity of current CMB missions and will probe increasingly larger scales, providing a unique opportunity to test our models of physics, gravity and the origin of large-scale structure.
 
The usefulness of the information we can extract from LSS is limited by the accuracy of our model predictions. Beyond the power-spectrum, one of the key observables is the bispectrum, measuring the correlation of three points in the sky. The bispectrum is directly linked to the non-Gaussianity of the perturbations since the bispectrum of a purely Gaussian distribution vanishes. If we can remove the bispectrum generated during the later epochs of the Universe from the observations, we can measure the primordial non-Gaussianity and constrain our models of inflation. In this paper we focus not on the primordial bispectrum, but on the bispectrum generated during the evolution of the perturbations after inflation. Even for purely Gaussian initial conditions, the dynamics beyond linear order will generate an unavoidable non-Gaussianity that needs to be modelled in detail for the analysis of future large scale structure surveys. While relativistic computations of the bispectrum are available \cite{Villa:2015ppa} that improve our understanding on large scales, the nonlinear impact of the radiation-dominated epoch on the Cold Dark Matter (CDM) distribution has yet to be fully understood. For the CMB bispectrum, it was shown that scattering interactions in the early photon-baryon plasma contribute to an intrinsic bispectrum which is just below observational limits from ESA's Planck satellite~\cite{Pitrou:2010sn,Huang:2012ub,Huang:2013qua,Su:2012gt,Su:2014tga,Pettinari:2013he, Pettinari:2014iha}. Their signature is then gravitationally imprinted  in the dark matter distribution as well, generating an \emph{intrinsic matter bispectrum}.

Although it originates in the radiation-dominated epoch, the intrinsic matter bispectrum affects the subsequent evolution of the CDM distribution.
A detailed understanding of this bispectrum is needed to correctly interpret results from future surveys such as Euclid \cite{laureijs2011euclid} and SKA~\cite{Maartens:2015mra}. In this paper we address this important issue by computing the CDM intrinsic bispectrum up to second order in the cosmological perturbations for the first time, both numerically using the code \SONG{}, and analytically in the squeezed limit. 

This paper is organised in the following way. In section~\ref{sec:Nperturbation} we review the bispectrum in Newtonian perturbation theory. In section~\ref{sec:GRperturbation} we generalise the Newtonian approach by including GR corrections, and we discuss the emergence of an intrinsic bispectrum in the dark matter perturbations during the epoch of radiation domination. In section~\ref{sec:squeezed} we perform an analytical computation of the CDM bispectrum in the squeezed-limit, consistently including for the first time the impact of radiation and baryons. Finally, in section~\ref{sec:results} we employ the code \SONG{} to compute the full second-order bispectrum including GR and radiation for all shapes. We compare these results against our analytic formulae, demonstrating the relative importance of the various contributions.

\subsection{Notation}

The evolution of matter is described by a set of partial differential equations. We will perform most computations in Fourier space instead of real space, leaving us with systems of ordinary differential equations instead. 
Our Fourier convention and notation is
\begin{align}
f(\vek{k}) = \Fourier{k}{f(\vek{x})} &= \int \di \vek{x} e^{-i\vek{k}\cdot\vek{x}} f(\vek{x}) \label{eq:FourierDef}.
\end{align}
The convolution theorem takes the simple form
\begin{equation}
\Fourier{k}{f(\vek{x}) g(\vek{x})} = \Kernel{\Fourier{k_1}{f(\vek{x})} \Fourier{k_2}{g(\vek{x})}}, \label{eq:convolution}
\end{equation}
where we have defined the convolution operator
\begin{align}
\Kernel{f(\vek{k}_1,\vek{k}_2) } &\equiv \int \frac{\di \vek{k}_1 \di\vek{k}_2}{(2\pi)^3} f(\vek{k}_1,\vek{k}_2) \delta^D(\vek{k}-\vek{k}_1-\vek{k}_2)\label{eq:KernelDef}
\end{align}
and $\delta^D(\vek{p})$ is the Dirac delta function.

We will use perturbation theory to describe the evolution of density inhomogeneities, expanding the small initial perturbations up to second order
\begin{equation}
\delta \simeq \delta^{(1)} + \frac{1}{2} \delta^{(2)} + \cdots, \label{eq:expansion}
\end{equation}
where first-order density perturbations describe a Gaussian random field. Non-linear interactions inevitably generate a non-Gaussian distribution at second and higher orders. Second-order density perturbations in Fourier space can be related to a convolution of the first-order perturbations
\begin{equation}
 \label{generalKernel}
\frac{1}{2} \delta^{(2)} (\vek{k}_3) = \mathcal{C}_{\vec{k}_3} \left\{  \mathcal{K}({k}_1,{k}_2,{k}_3) \delta^{(1)}(\vek{k}_1) \delta^{(1)}(\vek{k}_2) \right\} 
 \,,
\end{equation}
where the form of the kernel, $\mathcal{K}({k}_1,{k}_2,{k}_3)$, differs between Newtonian gravity and general relativity, and between various relativistic gauges.

In the following we will always write the kernel as a function of three wavenumbers $k_1$, $k_2$ and $k_3$. Although the explicit form of the kernel may include the cosine of the angle between two wavevectors, $\hat{\vek{k}}_1\cdot \hat{\vek{k}}_2$, this can always be expressed in terms of $k_1$, $k_2$ and the length of the vector $|\vek{k}_1+\vek{k}_2|$ through the relation
\begin{equation}
\hat{\vek{k}}_1\cdot \hat{\vek{k}}_2 = \frac{|\vek{k}_1+\vek{k}_2|^2-k_1^2-k_2^2}{2k_1k_2} \,,
\end{equation}
and translation invariance ensures that we are only ever interested in the value of the kernel when $|\vek{k}_1+\vek{k}_2|=k_3$. Since $\vek{k}_1$ and $\vec{k}_2$ are the integration variables, it is always possible to symmetrise the kernel with respect to the exchange of $k_1$ and $k_2$. Using this symmetry, without loss of generality we can consider only the case with $k_2 \leq k_1$. 

A Gaussian random field is described completely by the power spectrum in Fourier space
\begin{equation}
 \label{def:powerspectrum}
\langle \delta(\vek{k}_1) \delta(\vek{k}_2) \rangle = (2\pi)^3 P(k_1) \delta^D(\vek{k}_1+\vek{k}_2) \,.
\end{equation}
Second-order corrections lead to a non-vanishing bispectrum, defined by the 3-point function in Fourier space
\begin{equation}
 \label{def:bispectrum}
\langle \delta(\vek{k}_1) \delta(\vek{k}_2) \delta(\vek{k}_3) \rangle = (2\pi)^3 B({k}_1,{k}_2,{k}_3) \delta^D(\vek{k}_1+\vek{k}_2+\vek{k}_3) \,.
\end{equation}
Note that we are assuming an isotropic distribution in which case the power spectrum and bispectrum are functions only of the wavenumbers, $k_i=|\vek{k}_i|$, and independent from the direction of the wavevectors, $\hat{\vek{k}}_i$.

Substituting~\eqref{eq:expansion} and~(\ref{generalKernel})  into~(\ref{def:bispectrum}), and using Wick's theorem and~\eqref{def:powerspectrum}, gives the leading-order bispectrum for Gaussian initial conditions as
\begin{equation}
\label{redbispectrum}
B({k}_1,{k}_2,{k}_3) = 2 \left\{ P(k_1) P(k_2) \mathcal{K}({k}_1,{k}_2,{k}_3) + 2\ {\rm perms} \right\} \,.
\end{equation}

Thus in what follows we focus on the general relativistic kernel $\mathcal{K}({k}_1,{k}_2,{k}_3)$ obtained in different gauges and compare this with the kernel found in Newtonian theory.  
Among various configurations of the bispectrum, we study the following limits explicitly in this paper:
\begin{itemize}
	\item squeezed configurations: $k_2 \ll k_1 \sim k_3$.
	\item equilateral configurations: $k_1 \sim k_2 \sim k_3$.
	\item folded configurations: $k_1 + k_2 \sim k_3$. 
\end{itemize}

We denote the choice of gauge for the density perturbations by subscript $P$ for Poisson gauge and $t$ for total matter gauge.  Poisson gauge at first order is often called conformal Newtonian gauge or longitudinal gauge~\cite{Malik:2008im}. 
The line element (including only scalar perturbations) is given by
\begin{equation}
ds^2 = a(\tau)^2 \Big(   
- (1 + 2 \Psi) d\tau^2 + (1+ 2 \Phi) \delta_{ij} dx^i dx^j
\Big). 
\label{firstPoisson}
\end{equation}
Our $\Phi$ and $\Psi$ coincide with the notation of Kodama \& Sasaki~\cite{Kodama:1985bj}. In terms of $\phi^\text{MB}$ and $\psi^\text{MB}$ of Ma \& Bertschinger~\cite{Ma:1995ey} we have $\Phi=-\phi^\text{MB}$ and $\Psi=\psi^\text{MB}$. We will follow~\cite{Creminelli:2011sq} in defining the comoving curvature perturbation $\zeta$ at first order as
\begin{equation}
\zeta = \Phi - \frac{2}{3} \frac{1}{1+w} \left[ \Psi - \frac{a}{\dot{a}} \dot{\Phi} \right],
\label{zeta}
\end{equation}
where the dot denotes the derivative with respect to conformal time. In the analytic part of the paper we will assume vanishing anisotropic stress in which case $\Psi = -\Phi$ at first order. 

\section{Newtonian perturbation theory}\label{sec:Nperturbation}

\subsection{Fluid equations in comoving coordinates}

Before we present the GR results, let us briefly review some of the basic results from Newtonian perturbation theory. We use the Friedmann equations
\begin{align}
\Hc^2 &= H_0^2\left(\Omega_m a^{-1} + \Omega_\Lambda a^2\right),\\
\Hcdot &= \Hc^2 - \frac{3}{2} \frac{\HcO}{a},
\end{align}
where $H_0$ is the present-day Hubble constant, $\Hc\equiv \frac{\dot{a}}{a}$ is the conformal Hubble factor and a dot denotes a derivative with respect to conformal time. 
We consider analytic solutions in a flat universe $\Omega_m+\Omega_\Lambda=1$ with pressureless dust and a cosmological constant. 
The limit $\Omega_m\to1$ corresponds to a matter-dominated or Einstein-de Sitter (EdS) cosmology.
In our notation, $\Omega_m$ and $\Omega_\Lambda$ are the density fractions \emph{today} and are time independent. 

We define comoving coordinates and velocities $\vec{x}$ and $\vec{u}$ which are related to the physical coordinates and velocities $\vec{r}$ and $\vec{v}$ by
\begin{equation}
\vec{x} \equiv \frac{\vec{r}}{a(\tau)} \,,
\qquad
\vec{u} = \dot{\vec{x}} = \vec{v} - \Hc \vec{x} 
\, .
\label{eq:vpeculiar}
\end{equation}

The Newtonian dark matter density contrast $\delta(\tau,\vec{x})$ and the divergence of the peculiar velocity flow $\theta(\tau,\vec{x})\equiv \partial_i v^i$ satisfy two nonlinear equations,
\begin{align}
\dot{\delta} &= -\partial_j \left[ (1+\delta) \partial_j \nabla^{-2} \theta \right] \label{eq:deltanonlinear} \\
\dot{\theta} &= - \Hc \theta - \partial_i \partial_j \nabla^{-2}\theta \partial_j \partial_i \nabla^{-2}\theta - \partial_j \nabla^{-2} \theta \partial_j \theta - \frac{3}{2} \frac{\HcO}{a} \delta, \label{eq:thetanonlinear}
\end{align}
where $\partial_i \equiv \diffpar{}{x^i}$. Although these equations are well known (see for example~\cite{Bernardeau:2001qr}), we provide a derivation in appendix~\ref{sec:NewtonianBoltzmann} for completeness. 

\subsection{Cosmological perturbation theory}

We solve the dark matter equations of motion perturbatively. At first order, equations (\ref{eq:deltanonlinear}) and (\ref{eq:thetanonlinear}) become
\begin{align}
\dot{\delta}^{(1)}  &= - \theta^{(1)}, \\
\dot{\theta}^{(1)} &=-\Hc \theta^{(1)} - \frac{3}{2} \frac{\HcO}{a} \delta^{(1)}.
\end{align}
These can be combined to a single second-order equation for $\delta^{(1)}$,
\begin{equation}\label{eq:delone}
\ddot{\delta}^{(1)} + \Hc \dot{\delta}^{(1)} =  \frac{3}{2} \frac{\HcO}{a} \delta^{(1)}.
\end{equation}
Equation~\eqref{eq:delone} does not depend on spatial coordinates explicitly, so we can solve it by the product ansatz 
$\delta^{(1)}(\tau,\vec{x}) \equiv D(\tau) \deltil(\vec{x})$. This leads to an ordinary differential equation for the linear growth function $D(\tau)$:
\begin{equation}\label{eq:Dequation}
\ddot{D} + \Hc \dot{D} - \frac{3}{2} \frac{\HcO}{a}D = 0\;.
\end{equation}
Being a second-order differential equation, we generally find two solutions that are set by the initial conditions. One of these modes is sub-dominant and can typically be neglected; therefore, in the remainder of the paper $D$ refers to the fastest growing mode only. 
From the differential equation for $\delta^{(1)}$ we then find
\begin{equation}
\theta^{(1)} = -\dot{D} \deltil(\vec{x}) = -\frac{\dot{D}}{D} \delta^{(1)}.
\end{equation}

The equations for the second-order perturbations are more complicated. {}From equations (\ref{eq:deltanonlinear}) and (\ref{eq:thetanonlinear}) we have
\begin{align}
\dot{\delta}^{(2)} &= - 2\partial_j \nabla^{-2} \theta^{(1)} \partial_j \delta^{(1)} - 2\delta^{(1)} \theta^{(1)} - \theta^{(2)}, \\
\dot{\theta}^{(2)} &= - \Hc  \theta^{(2)} - 2\left(\partial_i \partial_j \nabla^{-2} \theta^{(1)}\partial_i \partial_j \nabla^{-2} \theta^{(1)}\right) - 2\partial_j \nabla^{-2} \theta^{(1)} \partial_j \theta^{(1)} - \frac{3}{2} \frac{\HcO}{a} \delta^{(2)}.
\end{align}
We can rewrite these equations by using the first-order solutions:
\begin{align}
\dot{\delta}^{(2)} &= 2D\dot{D} \left[ \partial_j \nabla^{-2}\deltil \partial_j \deltil + \deltil^2\right] - \theta^{(2)}, \\
\dot{\theta}^{(2)} &= - \Hc  \theta^{(2)} - \frac{3}{2} \frac{\HcO}{a} \delta^{(2)} - 2\dot{D}^2 \left[ \partial_i \partial_j \nabla^{-2}  \deltil \partial_i \partial_j \nabla^{-2}  \deltil + \partial_j \nabla^{-2}  \deltil \partial_j \deltil \right] .
\end{align}
Again we combine the equations to a single second-order equation for $\delta^{(2)}$:
\begin{align}\nonumber
 \ddot{\delta}^{(2)} + \Hc \dot{\delta}^{(2)} &=2 \left(\Hc \dot{D} D + \ddot{D} D + \dot{D}^2 \right)  \left[ \partial_j \nabla^{-2}\deltil \partial_j \deltil + \deltil^2\right]+ \frac{3}{2} \frac{\HcO}{a} \delta^{(2)} + \\
 &\quad 
 + 2\dot{D}^2 \left[ \partial_i \partial_j \nabla^{-2}  \deltil \partial_i \partial_j \nabla^{-2}  \deltil + \partial_j \nabla^{-2}  \deltil \partial_j \deltil \right] . \label{eq:Newtonian2nd}
\end{align}
At second order the structure explicitly depends on the spatial coordinates and a product ansatz is no longer possible. However, using the spatial functions
\begin{align}
B_1(\vec{x}) &\equiv 2 \partial_j \nabla^{-2}\deltil \partial_j \deltil, & B_2(\vec{x}) &\equiv 2 \deltil^2, & B_3(\vec{x}) &\equiv  2\partial_i \partial_j \nabla^{-2}  \deltil \partial_i \partial_j \nabla^{-2}  \deltil ,
\end{align}
we can rewrite the differential equation as
\begin{align}
 \ddot{\delta}^{(2)} + \Hc \dot{\delta}^{(2)} - \frac{3}{2} \frac{\HcO}{a} \delta^{(2)}  
\nonumber
&= \left(\Hc \dot{D} D + \ddot{D} D + 2\dot{D}^2 \right) \left(B_1 +\frac{B_2+B_3}{2}\right) +\\
& \quad
 + \left(\Hc \dot{D} D + \ddot{D} D \right)  \frac{B_2-B_3}{2}.
 \label{eq:newtondeltasecond} 
 \end{align}
Now we can obtain a particular solution to this equation by making the ansatz
\begin{align}\nonumber
\delta^{(2)} &= D^2 \left( b_1(\tau) B_1(\vec{x}) +  b_{+}(\tau) \frac{B_2(\vec{x})+B_3(\vec{x})}{2} + b_{-}(\tau) \frac{B_2(\vec{x})-B_3(\vec{x})}{2} \right)\\
&\equiv D^2 \left( b_1(\tau) B_1(\vec{x}) +  b_{+}(\tau)B_{+}(\vec{x})+ b_{-}(\tau) B_{-}(\vec{x}) \right). 
\end{align}
Inserting this ansatz in equation~\eqref{eq:newtondeltasecond} gives
\begin{align}
 \ddot{\delta}^{(2)} &+ \Hc \dot{\delta}^{(2)} - \frac{3}{2} \frac{\HcO}{a} \delta^{(2)} = \nonumber \\
 &\sum_{i\in \{1,+,-\}} \left\{ \left(\Hc \dot{D} D + \ddot{D} D + 2\dot{D}^2 \right) b_i + \left(4 D \dot{D} + \Hc D^2\right) \dot{b}_i + 2D^2 \ddot{b}_i \right\} B_i,
 \end{align}
where we have used equation~\eqref{eq:Dequation}. By comparing this result with equation~\eqref{eq:newtondeltasecond} we immediately find two of the three coefficients $b_1=b_{+} = 1$ and the remaining function $b_{-}$ satisfies the differential equation
 \begin{equation}
 \left(\Hc \dot{D} D + \ddot{D} D + 2\dot{D}^2 \right) b_{-} + \left(4 D \dot{D} + \Hc D^2\right) \dot{b}_{-} + 2D^2 \ddot{b}_{-} 
 = \frac{3}{2} \HcO \frac{D^2}{a}.
 \end{equation}
We define the second-order growth function $F\equiv D^2 b_{-}$ and find an equation for $F$ which is very similar to the equation for the linear growth function $D(\tau)$ equation~(\ref{eq:Dequation}):
 \begin{equation}\label{eq:Fequation}
 \ddot{F} + \Hc \dot{F} = \frac{3}{2} \frac{\HcO}{a} \left(F+D^2 \right).
 \end{equation}
 The particular solution of the second-order Newtonian density contrast is finally given by~\cite{Matsubara:1995kq}
 \begin{align}
 \delta^{(2)} &= 2\partial_j \nabla^{-2}\delone \partial_j \delone +  \left(1+\frac{F}{D^2}\right)  \delone \delone +  \left(1-\frac{F}{D^2}\right) \partial_i \partial_j \nabla^{-2}  \delone \partial_i \partial_j \nabla^{-2}  \delone \label{eq:delta2real}
 \end{align}
 
By applying the Fourier transform~\eqref{eq:FourierDef}, we can write the result in Fourier space using the convolution operator \eqref{eq:KernelDef}. The result is
 \begin{align}
 \frac{1}{2}\delta^{(2)} (\vek{k}) &= \Kernel{\mathcal{K}_N (k_1,k_2,k) \delta^{(1)} \left( \vek{k}_1  \right) \delta^{(1)} \left( \vek{k}_2 \right) }, \label{eq:delta2N} \\
 \mathcal{K}_N (k_1,k_2,k) &\equiv \left( \beta_N - \alpha_N \right) + \frac{\beta_N}{2} \coskk \left( \frac{k_2}{k_1} + \frac{k_1}{k_2} \right)  + \alpha_N \left(\coskk \right)^2,  \label{eq:kernelN}
 \end{align}
where the dimensionless coefficients, adopting the notation of Ref.~\cite{Fitzpatrick:2009ci}, are given by
 \begin{equation} 
 \alpha_N = \frac{7-3v}{14}, \quad
 \beta_N = 1,
 \end{equation}
We have defined the ratio of the second-order and first-order growth functions, $v \equiv 7 F/3 D^2$ (see appendix~\ref{sec:growth}). In the matter-dominated (EdS) limit we have $v\to1$ and $\alpha_N\to2/7$.
 
The full solution of the second-order perturbation equation~(\ref{eq:Newtonian2nd}) is given by the particular solution~(\ref{eq:delta2real}) plus the homogeneous solution that obeys the same equation as the linear perturbation equation~(\ref{eq:delone}). The coefficients of the homogeneous solution must be fixed by the initial conditions. However in Newtonian theory we have only the linear Poisson constraint relating the initial density perturbation to the initial potential. Therefore 
for a Gaussian primordial potential the homogeneous solution must be set to zero for consistency beyond first order leaving only the particular solution~(\ref{eq:delta2real}).

Second-order perturbations are thus vanishing initially in the Newtonian theory for Gaussian initial conditions but nonlinear density perturbations are generated from the source terms in equation~(\ref{eq:Newtonian2nd}), quadratic in the first-order perturbations.
As we shall see, this is no longer true in general relativity where non-linear constraint equations require non-vanishing initial density perturbations at second and higher orders~\cite{Bartolo:2005xa,Bruni:2013qta,Bruni:2014xma,Uggla:2013kya,Villa:2015ppa}.
 
\section{General relativistic perturbation theory}\label{sec:GRperturbation}

While we expect Newtonian theory to provide a good description of the growth of structure on small scales, large scales that are close to the size of the horizon cannot be described without using a relativistic framework. The spatial and temporal evolution of structure in general relativity (GR) depends on the choice of space and time coordinates used to describe the evolution. In this section we analyse the impact of GR for two different gauge choices. 

\subsection{Total matter gauge}\label{sec:NewtonianComparison}

The evolution of the second-order density contrast in General Relativity is most easily compared to the Newtonian result, \eqref{eq:delta2N}, in the total matter gauge~\cite{Noh:2004bc,Yoo:2014vta,Bertacca:2015mca}.  This gauge shares the same spatial coordinates as the Poisson gauge and the same time-slicing as the synchronous-comoving gauge~\cite{Liddle:2000cg,Malik:2008im}. For example, the relativistic constraint equation for the first-order Bardeen potential has the standard form for the Newtonian Poisson equation when written in terms of the total matter gauge density contrast:
\begin{equation}
\nabla^2\Phi = -\frac{3}{2} \frac{\HcO}{a} \delta^{(1)}_t \,. \label{eq:tPoissonEquation}
\end{equation}
Also, in the total matter gauge the second order equation of motion for the density contrast is identical to the Newtonian one \cite{Noh:2004bc}, and therefore has the same particular solution, equation~\eqref{eq:delta2real}. 

However, an important difference in GR at second order is that one has nonlinear constraint equations in addition to the dynamical equations. While in Newtonian theory the second-order initial density perturbations may be set to vanish, this is not consistent with the GR constraints for a Gaussian primordial potential. Instead, we must impose a non-vanishing initial second-order density perturbation, quadratic in the initial first-order perturbations, that can be evaluated in simple large-scale limits \cite{Bruni:2014xma}, on super-horizon scales at the beginning of the matter-dominated era. 

Combining the particular solution~\eqref{eq:delta2real} and the homogeneous solution obtained from the initial GR constraints in the total matter gauge, we obtain~\cite{Bartolo:2010rw,Bruni:2013qta,Villa:2015ppa}
\begin{align}\nonumber
\delta_t^{(2)} &= 
 \label{eq:delta2tomreal}
2\partial_j \nabla^{-2}\delone \partial_j \delone +  \left(1+\frac{F}{D^2}\right)  \delone \delone +  \left(1-\frac{F}{D^2}\right) \partial_i \partial_j \nabla^{-2}  \delone \partial_i \partial_j \nabla^{-2} 
\delone
\\
& \quad
+ \frac{10 \Omega_m H_0^2}{D} \frac{D_\text{ini}}{a_\text{ini}}  \left[ -\frac{1}{4} \partial_i \nabla^{-2} \delone \partial_i \nabla^{-2} \delone+ \delone \nabla^{-2} \delone \right],
\end{align}
where $D_\text{ini}$ and $a_\text{ini}$ are the linear growth function and the scale factor at some initial time respectively. 
The first line is exactly the Newtonian particular solution, while the second line is the correction from GR modifying the initial perturbations. At late times the solution is dominated by the Newtonian solution, driven by source terms, and on a given scale it will eventually lose information about the initial conditions. This is rapidly the case on the small scales, leaving the GR corrections important only on large scales.

In Fourier space, the second-order GR density in the total matter gauge can be written in a way similar to the second-order Newtonian density contrast equation~(\ref{eq:delta2N}):
\begin{align}
\frac{1}{2}\delta_t^{(2)} &= \Kernel{\mathcal{K}_t(k_1,k_2,k)  \delta_t^{(1)} \left( \vek{k}_1 \right) \delta_t^{(1)} \left( \vek{k}_2 \right) },  \label{eq:kernelt} \\
\mathcal{K}_t(k_1,k_2,k) &\equiv \left( \beta_t - \alpha_t \right) + \frac{\beta_t}{2} \coskk \left( \frac{k_2}{k_1} + \frac{k_1}{k_2} \right) + \alpha_t \left(\coskk \right)^2 + \gamma_t\left( \frac{k_1}{k_2} - \frac{k_2}{k_1} \right)^2, \nonumber
\end{align}
where 
\begin{align} \nonumber
\alpha_t &= \frac{7-3v}{14} +  \left(f + \frac{3}{2} u \right) \frac{\Hc^2}{k^2}, \\ \nonumber
\beta_t &= 1 - 3 \left(f + \frac{3}{2} u \right) \frac{\Hc^2}{k^2}, \\
\gamma_t &= - \left(f + \frac{3}{2} u \right) \frac{\Hc^2}{k^2} \,.
\end{align}
Here we defined the linear growth rate $f\equiv\dot{D}/D {\cal H}$, the comoving matter fraction $u \equiv 1/(1 +  a^3 \Omega_{\Lambda}/\Omega_M)$ and we used the relation
\begin{equation}
\frac{D_\text{ini}}{a_\text{ini}} = \frac{2}{5} \frac{D{\cal H}^2}{\Omega_m H_0^2}
\left(f + \frac{3}{2} u \right),
\end{equation}
to eliminate the dependence on the normalisation of $D$.

Compared with the Newtonian kernel (\ref{eq:kernelN}), we see that incorporating GR 
introduces corrections in $\alpha_t$ and $\beta_t$ proportional to $\Hc^2/k^2$ and also gives an entirely new $\gamma_t$ term that is again proportional to  $\Hc^2/k^2$. This implies that the GR corrections become important on scales comparable to the horizon. The new term proportional to $\gamma_t$ dominates the kernel in the squeezed limit where $k_2 \ll k_1 \sim k$.
This term is absent in the Newtonian treatment and this aspect makes the squeezed limit an important potential testing ground for relativistic effects. 
We will discuss this limit in detail in section~\ref{sec:squeezed}. 

\subsection{Poisson gauge}

We are particularly interested in the second-order CDM density in Poisson gauge, since the numerical Einstein-Boltzmann code \SONG{} works in this gauge. \SONG{}  uses Poisson gauge because the photon scattering term is most straightforward in this gauge~\cite{pettinari:2015a}. However, unlike in the total matter gauge, the constraint relating the potential, $\Phi$, to the first-order density contrast in this gauge, $\delta_P^{(1)}$, also receives a GR correction at first order
\begin{equation}
\left[ \nabla^2 - 3f {\cal H}^2 \right] \Phi = - \frac{3}{2} \frac{\HcO}{a} \delta^{(1)}_P \,. \label{eq:pPoissonEquation}
\end{equation}

The expression for the second-order density contrast in Poisson gauge is more complicated than the equivalent expression in total matter gauge~\eqref{eq:delta2tomreal}. It was recently derived in $\Lambda$CDM by Villa and Rampf in real space~\cite{Villa:2015ppa}
(see also \cite{Uggla:2013kya}). We Fourier transform the second-order contribution to the Poisson gauge density $\delta_P^{(2)}$, equation (5.54) in~\cite{Villa:2015ppa}, by using our Fourier dictionary in appendix~\ref{sec:dict}:
\begin{align} 
\frac{1}{2}\delta_P^{(2)} (\vek{k}) &= \Kernel{\mathcal{K}_{P}(k_1,k_2,k)  \delta_P^{(1)} \left( \vek{k}_1 \right) \delta_P^{(1)} \left( \vek{k}_2 \right) },  \label{eq:delta2p} \\
\mathcal{K}_{P,\text{VR}} &\equiv \frac{
 \left( \beta_{P,\text{VR}} - \alpha_{P,\text{VR}} \right) + \frac{\beta_{P,\text{VR}}}{2} \coskk \left( \frac{k_2}{k_1} + \frac{k_1}{k_2} \right) +  \alpha_{P,\text{VR}} \left(\coskk \right)^2 +   \gamma_{P,\text{VR}} \left( \frac{k_1}{k_2} - \frac{k_2}{k_1} \right)^2
 }{\left(1+3f\frac{\Hc^2}{k_1^2} \right) \left(1+3f\frac{\Hc^2}{k_2^2} \right)}
 . \nonumber
\end{align}
We have simplified the Poisson gauge kernel by pulling out the two factors $\left(1+3f\Hc^2/k_1^2\right)^{-1}$ and $\left(1+3f\Hc^2/k_2^2\right)^{-1}$. This effectively transforms the first-order Poisson gauge densities into total matter gauge densities as one can verify by comparing equation~\eqref{eq:tPoissonEquation} and equation~\eqref{eq:pPoissonEquation} in Fourier space. The coefficients in the kernel are given by
\begin{align} \nonumber
\alpha_{P,\text{VR}} &= \frac{7-3v}{14}+\left(4f+\frac{3}{2}u-\frac{9}{7}w\right) \frac{\Hc^2}{k^2}+\left(18 f^2+9f^2u-\frac{9}{2} f u\right) \frac{\Hc^4}{k^4}, \\ \nonumber
\beta_{P,\text{VR}} &= 1 +\left(-2 f^2+6f-\frac{9}{2}u \right) \frac{\Hc^2}{k^2}+\left( 36 f^2+18 f^2 u \right) \frac{\Hc^4}{k^4}, \\
\gamma_{P,\text{VR}} &= \frac{1}{2} \left(-f^2+f-3 u\right) \frac{\Hc^2}{k^2}+\frac{1}{4} \left(18f^2+9(f^2-f) u \right) \frac{\Hc^4}{k^4}, \label{eq:gammap}
\end{align}
and $w$ is the second-order growth rate, $w\equiv 7 \dot{F}/6 \Hc D^2$.  
The $\Lambda CDM$ growth functions $f,u,v,w$ satisfy $f=u=v=w=1$ in the EdS limit and 
equation~(\ref{eq:gammap}) reduces to those obtained in Ref.~\cite{Fitzpatrick:2009ci} in this limit:
\begin{align} \nonumber
\alpha_\text{P,EdS} &= \frac{2}{7} + \frac{59 \Hc^2}{14 k^2}
+ \frac{45\Hc^4}{2 k^4}, \\ \nonumber
\beta_\text{P,EdS}  &= 1 - \frac{\Hc^2}{2 k^2} + \frac{54 \Hc^4}{k^4}, \\
\gamma_\text{P,EdS} &=- \frac{3\Hc^2}{2 k^2}+ \frac{9\Hc^4}{2k^4}. \label{eq:gammaFSZ}
\end{align}
Compared with the kernel in the total matter gauge (\ref{eq:kernelt}), there are additional corrections from GR in the Poisson gauge that are important when $\Hc^2/k^2$ becomes large. These additional corrections arise due to a different time slicing in the two gauges.

\subsection{Early Universe radiation}

\begin{figure}[tb]
\begin{center}
\includegraphics[width=\textwidth]{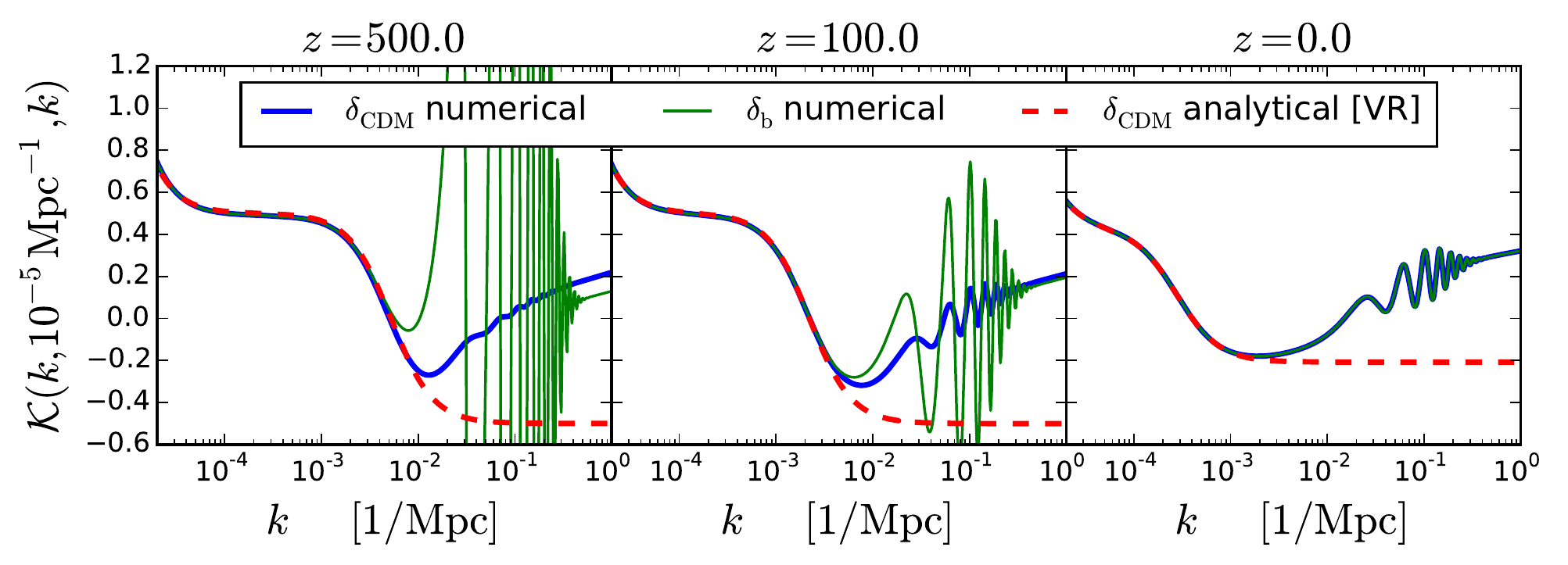}
\end{center}
\caption{The second-order kernel obtained at three different redshifts from the numerical Einstein-Boltzmann code \SONG{} for cold dark matter (thick blue line) and baryons (thin green line), compared with the analytic kernel~\eqref{eq:delta2p} obtained by Villa and Rampf~\cite{Villa:2015ppa} (dashed red line), for squeezed configurations with the longest wavelength mode $k_2=10^{-5}$Mpc$^{-1}\ll k_1\simeq k$.}
\label{fig:squeezedproblem}
\end{figure}

The calculations in the preceding sections implicitly assume that all modes enter the horizon when the Universe is matter dominated. However, a wide range of modes ($k > k_\text{eq}$, with $k_\text{eq}\simeq0.01~\text{Mpc}^{-1}$) enter the horizon before matter-radiation equality. These modes are affected by the complicated early Universe physics which requires us to solve the full Einstein-Boltzmann system, including the effects of radiation and neutrinos. We can simplify the problem if we restrict our attention to finding solutions valid in only matter-domination, but with modified initial conditions after matter-radiation equality. In this case the Newtonian evolution equations, (\ref{eq:delone}) and (\ref{eq:Newtonian2nd}), are still valid, but the preceding phase of radiation domination modifies the initial conditions at the start of the matter era and thus enters the homogeneous part of the solution, similar to the GR corrections. 

Modes entering the horizon during radiation domination are subject to a variety of scattering effects \cite{pitrou:2010a, Pettinari:2013he, Pettinari:2014iha, Huang:2012ub, Huang:2013qua,Su:2012gt, Su:2014tga} that contribute to the emergence of an intrinsic bispectrum in the CMB and in the baryon distribution, which is then gravitationally imprinted in the dark matter distribution. 

To demonstrate the impact of the intrinsic bispectrum from radiation domination on the dark matter distribution, we show in \fref{fig:squeezedproblem} the  analytic kernel in Poisson gauge given in equation~\eqref{eq:delta2p} with the one extracted from \SONG{} solving the full equations for a squeezed configuration with $k_2=10^{-5}$Mpc$^{-1}$.  The modifications to the initial conditions are most relevant on the large scales, as discussed for the relativistic corrections.  However, to see the effect of the radiation physics we need a mode small enough so it enters the horizon during radiation domination. The squeezed limit combines a long and a short mode so it is a natural configuration for studying the radiation correction. As expected, we find that the analytic solution breaks down when the short-wavelength wavenumber exceeds $k_\text{eq}$.

This effect is caused by two major contributions. First, as discussed above, the dark matter initial conditions are changed gravitationally due to the presence of radiation. \SONG{} also includes baryons which are more directly affected by radiation via Compton scattering. Using the same arguments, their initial distribution at the onset of matter domination is modified. While this does not directly translate into a change in the dark matter bispectrum at high redshift ($z\sim500$), the baryons will gravitationally attract the dark matter particles during matter domination, leaving a distinctive signature; the {\em baryon acoustic oscillations} of the dark matter bispectrum. The plot shows the imprint of these characteristic oscillations in addition to a smooth rise in the bispectrum related to solving general relativity in radiation domination. 

In the next section we shall derive a squeezed-limit approximation of the intrinsic bispectrum that, when combined with the preceding analytical prediction in \eqref{eq:gammap}, will yield a much better match with \SONG{}.

\section{Squeezed-limit approximation}\label{sec:squeezed}

As shown in figure~\ref{fig:squeezedproblem}, in the squeezed limit, $k_2 \ll k_1 \sim k$, the analytic formula (\ref{eq:gammap}) requires improvement to reproduce the results from \SONG{}. In this squeezed limit, it is possible to independently derive the second-order density contrast by extending the first-order perturbations based on the separate universe approach, including the missing radiation correction.

We focus on a patch in the Universe much smaller than the long-wavelength mode $k_2 \equiv k_l$ but large enough to contain the short mode $k_1 \equiv k_s$. The main idea of the separate universe approach~\cite{Salopek:1990jq,Wands:2000dp} is that the long-wavelength mode can be considered as a locally homogeneous background within this patch. It is then possible to define {\it local} coordinates in this patch where the long-wavelength mode is removed by local coordinate transformations from global coordinates. In these coordinates, the short wavelength mode evolves independently of the long-wavelength mode. The coupling between the long and short wavelength modes appears when we move back to global coordinates and this coupling generates the second-order density contrast in the squeezed limit in global coordinates~\cite{creminelli:2004a,Fitzpatrick:2009ci,Creminelli:2011sq,Lewis:2012tc,Creminelli:2013mca}.  

The coordinate transformation locally removing a long-wavelength comoving curvature perturbation $\zeta$ from the metric in Poisson gauge is given by~\cite{Weinberg:2003sw}: 
\begin{align}
\etilde &= \eta + \epsilon(\eta),\\
\xtilde^j &= x^j(1+\zeta),
\end{align}
where $\epsilon(\eta)$ is 
\begin{equation}
\epsilon(\eta) = -\frac{\zeta}{a^2} \int_0^a \frac{a'}{\Hc(a')} da' \equiv E(\eta) \zeta,\label{eq:Eequation}
\end{equation}
in the limit of vanishing anisotropic stress. In EdS, we have $E(\eta)=-\eta/5$, while it is given by a hypergeometric function in $\Lambda$CDM as shown in appendix~\ref{sec:growth}. 

In the new coordinates the comoving curvature of the long mode is vanishing and we compute the matter over-density $\delta\rho(\etilde,\xtilde^i)$ on a patch which is small compared to the long mode. The full second-order solution in global coordinates is obtained by inverting the coordinate transformation, thereby adding the impact of the long mode on the short one. We relate the full matter over-density $\delta\rho(\eta,x^i)$ to $\delta\rho(\etilde,\xtilde^i)$ by Taylor expansion, discarding all terms of higher than second order in perturbations:
\begin{align} \nonumber
\delta\rho_P(\etilde,\xtilde^i) &= \delta\rho_P(\eta,x^i) + \frac{\partial \delta\rho_P}{\partial \eta}(\eta,x^i) (\etilde-\eta) + \frac{\partial \delta\rho_P}{\partial x^j} (\eta,x^i) (\xtilde^j-x^j), \\
&=  \bar{\rho} \delta_P(\eta,x^i) +  \bar{\rho}  \left(-3\Hc \delta_P(\eta,x^i) +\dot{\delta}_P(\eta,x^i) \right) E \zeta + \bar{\rho} \frac{\partial \delta_P}{\partial x^j} (\eta,x^i) x^j \zeta. \label{eq:squeezedpart1}
\end{align}
After dividing by the background density, $\bar{\rho}$, we identify the two terms as the second-order contribution to $\delta_P$,
\begin{equation}
\frac{1}{2} \delta^{(2)}_P(\eta,\vec{x}) = \left[ \left(-3\Hc \delta_P^{(1)}(\eta,\vec{x}) +\dot{\delta}_P^{(1)}(\eta,\vec{x}) \right)  E + \frac{\partial \delta_P^{(1)}}{\partial x^j} (\eta,\vec{x}) x^j \right]\zeta,
\end{equation}
which in Fourier space becomes
\begin{align}
\frac{1}{2} \delta^{(2)}_P(\eta,\vec{k}) &= \Kernel{\Fourier{\vec{k_s}}{\left(-3\Hc \delta_P^{(1)}(\eta,x) +\dot{\delta}_P^{(1)}(\eta,x) \right) E  + \frac{\partial \delta_P^{(1)}}{\partial x^j} (\eta,x) x^j} \zeta(\vec{k_l})}, \\
&= \Kernel{ \left[ \left(3-\frac{\dot{\delta}^{(1)}_P}{\Hc \delta^{(1)}_P} \right) \frac{1}{1+\frac{3u}{2f}} - \left(3 + \dlogpar{\delta^{(1)}_P}{k}\right) \right] \delta^{(1)}_P(\vec{k_s}) \zeta(\vec{k_l})},\label{eq:squeezedalmostdone}
\end{align}
where $\vec{k_s}$ is associated with the short mode density contrast in the small patch while $\vec{k_l}$ is associated with the long mode curvature perturbation. Here, we eliminate $\Hc E$ using the relation $\Hc E =  -(1+\frac{3u}{2f})^{-1}$ that we derive in appendix~\ref{sec:separate}.

From the Poisson equation~\eqref{eq:pPoissonEquation} we find
\begin{align}
\delta_P^{(1)}(\eta,\vek{k}) &= \frac{2}{3u} \left(\frac{k_s^2}{\Hc^2}+3f\right) \Phi(\eta,\vek{k}),\\
\dot{\delta}_P^{(1)}(\eta,\vek{k}) &=\frac{2}{3u} \Hc \left[f\frac{k^2}{\Hc^2}+\frac{9}{2} u (1-f) \right] \Phi(\eta,\vek{k}),
\end{align}
where we have used the differential equation for the growth function $D$ in the computation of the second identity.

We can write the Bardeen potential $\Phi$ as a transfer function $T(\eta,k)$ times a primordial random field $\Phi_{\vek{k}}$ (e.g., set by inflation)
\begin{equation}
\Phi(\eta,\vek{k}) = T(\eta,k) \Phi_{\vek{k}}.
\end{equation}
We then find
\begin{equation}
3 + \dlogpar{\delta^{(1)}_P}{k} = \frac{2}{1+3f\frac{\Hc^2}{k^2}} + 3 + \dlogpar{\Phi}{k} = \frac{2}{1+3f\frac{\Hc^2}{k^2}} + \dlogpar{T}{k} 
+ \dlogpar{(k^3 \Phi_{\vek{k}})}{k}.
\label{eq:dlogpar}
\end{equation}
The last term in~\eqref{eq:dlogpar} leads to 
primordial non-Gaussianity 
for adiabatic perturbations
which is proportional to $n_s-1$ in the squeezed limit, where $n_s-1$ is the tilt of the primordial power spectrum~\cite{Maldacena:2002vr,Creminelli:2004yq}. In this paper, since we are interested in the intrinsic matter bispectrum, we take $n_s = 1$ for simplicity and therefore neglect the last term. 

The relationship between $\zeta$ and $\Phi$ can be obtained by using equation~(\ref{zeta}) in the limit of vanishing anisotropic stress as
\begin{align}
\zeta &= \Phi + \frac{2}{3} \frac{1}{1+w} \left( \Phi + \frac{1}{\Hc} \dot{\Phi} \right) =\left(1 + \frac{2f}{3u} \right) \Phi
\end{align}
where we used $u=1+w$ and $\dot{\Phi}=\Hc (f-1)\Phi$. The latter formulae is most easily obtained by noting that equation~\eqref{eq:tPoissonEquation} implies $\Phi \propto D/a$. Inserting equation~\eqref{eq:dlogpar} into the second-order contribution, equation~\eqref{eq:squeezedalmostdone}, then gives
\begin{align}
&\frac{1}{2} \delta_P^{(2)}(\eta,\vec{k}_s) = \mathcal{C}_\vec{k} \Bigg\{ \bigg[ \left( f-f^2 -3u \right) \frac{\Hc^2}{k_s^2}   +9\left(f^2 - \frac{1}{2} fu (1-f)  \right) \frac{\Hc^4}{k_s^4}  \nonumber \\
&\qquad  -\left( f+\frac{3u}{2} \right)  \left(\frac{\Hc^2}{k_s^2}+3f \frac{\Hc^4}{k_s^4}\right) \dlogpar{T}{k} \bigg] \frac{1}{2} \left(\frac{k_s}{k_l} - \frac{k_l}{k_s}\right)^2  
\frac{\delta(\vec{k}_s)}{1+3f\frac{\Hc^2}{k_s^2}} \frac{\delta(\vec{k}_l)}{1+3f\frac{\Hc^2}{k_l^2}} \Bigg\},
\label{eq:2ndpoisson}
\end{align}
where we symmetrised the kernel by using $\frac{k_s^2}{k_l^2} \rightarrow \frac{1}{2} \left(\frac{k_s}{k_l} - \frac{k_l}{k_s}\right)^2$. After identifying $k_s=k_1=k$ and $k_l=k_2$, we see that the second-order result can be expressed as a kernel of the form $\mathcal{K}_P$ given in~(\ref{eq:delta2p}) with a coefficient $\gamma_P$ given by this squeezed limit
\begin{align}
\gamma_{P,\text{sq}} &= \frac{1}{2}\left[ f-f^2-3u \right] \frac{\Hc^2}{k^2} 
+\frac{9}{2} \left[f^2 - \frac{1}{2} f u (1-f)  \right] \frac{\Hc^4}{k^4}   \nonumber \\
&\qquad -\frac{1}{2} \left( f+\frac{3u}{2} \right)  \left[\frac{\Hc^2}{k^2}+3f \frac{\Hc^4}{k^4}\right] \dlogpar{T}{k} \label{eq:gamma_squeezed}. 
\end{align}
The first two terms in~\eqref{eq:gamma_squeezed} match $\gamma_{P,\text{VR}}$ in equation~\eqref{eq:gammap} exactly. The last term in~\eqref{eq:gamma_squeezed} vanishes for a scale-invariant transfer function, i.e., for a scale-invariant distribution of perturbations after the radiation epoch, but in general it accounts for the impact of the preceding phase of radiation domination on perturbations at the start of the matter era and the gravitational effect of baryons.

Note that in the squeezed limit we can neglect the contributions to the kernel $\mathcal{K}_P$ in \eqref{eq:delta2p} coming from the terms proportional to $\alpha_{P,\text{VR}}$ and $\beta_{P,\text{VR}}$, while in the opposite limit, for equilateral shapes, the contribution to the kernel from $\gamma_{P,\text{sq}}$ is zero. Therefore the correction we have computed in this section to $\gamma_{P,\text{VR}}$ can be included in the full kernel for any shape without spoiling the non-squeezed limit results. 

The technique presented here for calculating the second-order density kernel is limited to squeezed shapes. There will be additional corrections to the kernel from the preceding radiation dominated era for other shapes which should in principle be considered. However, in the next section we show that these are usually small and modifying $\gamma_{P,\text{sq}}$ alone greatly improves the relativistic dark matter kernel.

\section{Numerical computation}\label{sec:results}

In order to test the validity of our analytic approximations, we compare them with the CDM bispectrum kernel computed by \SONG{}, a second-order Boltzmann code that includes the effect of photons, baryons and neutrinos. \SONG{} is written in Poisson gauge so all results in this section will be in this gauge.

\subsection{The numerical code SONG}

\SONG{} was originally conceived to compute the effect of non-linear dynamics on CMB observables like the CMB intrinsic bispectrum \cite{Pettinari:2013he, Pettinari:2014iha} and the B-polarisation power spectrum \cite{fidler:2014a}. The late-time dark matter kernels presented are not central to these tasks, but the code has all the structure needed to compute them efficiently. All kernels (metric, CDM, baryon, photon and neutrinos) can be obtained on an average 8-core machine for a single Fourier mode in about a millisecond. This high computational speed allowed us to perform extensive analytical and numerical tests on the kernels produced by \SONG{} \cite{pettinari:2015a}.

\SONG{} computes the perturbation kernels by solving the Einstein-Boltzmann system of coupled differential equations at second order in the Poisson gauge. The relativistic species (photons and neutrinos) are evolved in full, considering both anisotropic stresses and higher moments; for the massive ones (baryons and cold dark matter) \SONG{} employs a fluid approximation whereby it retains only the density and velocity moments\footnote{Note that \SONG{}'s structure allows for a more complex treatment of massive species (e.g. including pressure), as it evolves them using a momentum-integrated Boltzmann hierarchy \cite{pettinari:2015a,lewis:2002b}.}. GR relativistic effects are naturally accounted for and so are the interactions they induce between between the various species. The scattering effects between photons and baryons are included up to second order in the cosmological perturbations \cite{beneke:2010a} and, before the time of recombination, up to first order in the tight-coupling approximation. For a detailed description of the equations, refer to Sec. 5.3 of Ref.~\cite{pettinari:2015a}.

The initial conditions are obtained by solving the system of equations analytically deep in the radiation era, assuming adiabatic perturbations and super-horizon modes. This procedure yields constant non-vanishing initial conditions for the Newtonian potentials $\Phi$ and $\Psi$ and for the densities of the matter species, in agreement with our arguments in Sec. \ref{sec:NewtonianComparison}. It is also assumed that the post-inflationary Universe is initially Gaussian, with a vanishing nonlinear parameter $f_{\text{NL}}$ for the $\zeta$ curvature perturbation. Again, a detailed derivation of the second-order initial conditions used in \SONG{} is presented in Sec. 5.4 of Ref.~\cite{pettinari:2015a}.

\SONG{}'s layout is inspired by the first-order Boltzmann code \CLASS{} \cite{lesgourgues:2011a, blas:2011a}. In particular, \SONG{} inherits from \CLASS{} its modular structure and a differential equation solver designed for stiff systems (\emph{ndf15}). 
Like \CLASS{}, \SONG{} is open-source and is available online at \url{https://github.com/coccoinomane/song} in prerelease form; all numerical results presented here can be reproduced in this way. We plan to release the 1.0 version of \SONG{} later in 2016.

\subsection{Comparison of the analytic results}
\begin{figure}[tb]
\begin{center}
\includegraphics[width=1.0\textwidth]{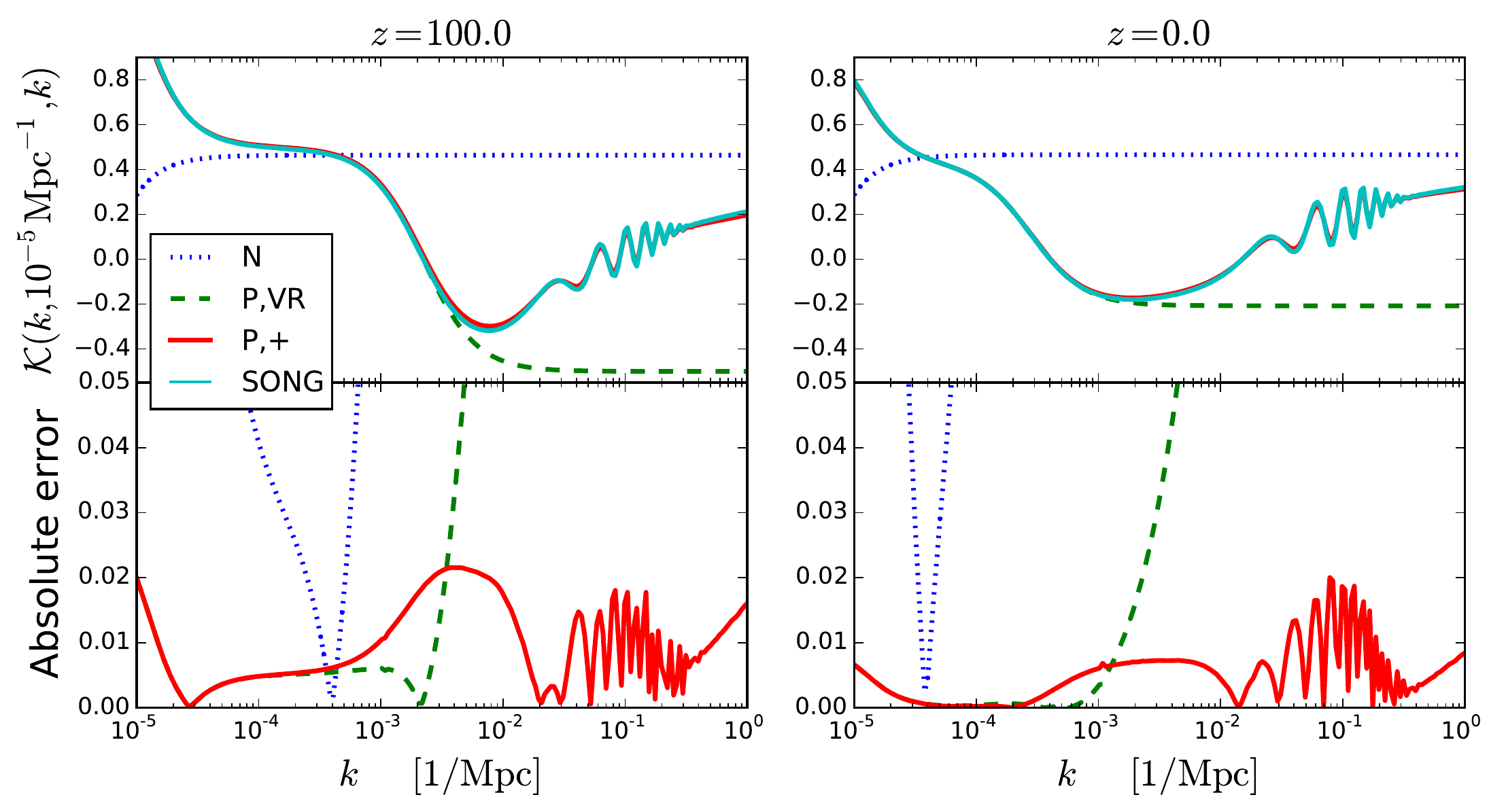}
\end{center}
\caption{The kernel computed by \SONG{} for a squeezed configuration compared to the analytic estimates at redshift $z=100$ and $z=0$. The squeezed limit correction derived in section~\ref{sec:squeezed} captures the effect of radiation in the initial conditions.}
\label{fig:simplesqueezed}
\end{figure} 
\begin{figure}[tb]
\begin{center}
\includegraphics[width=1.0\textwidth]{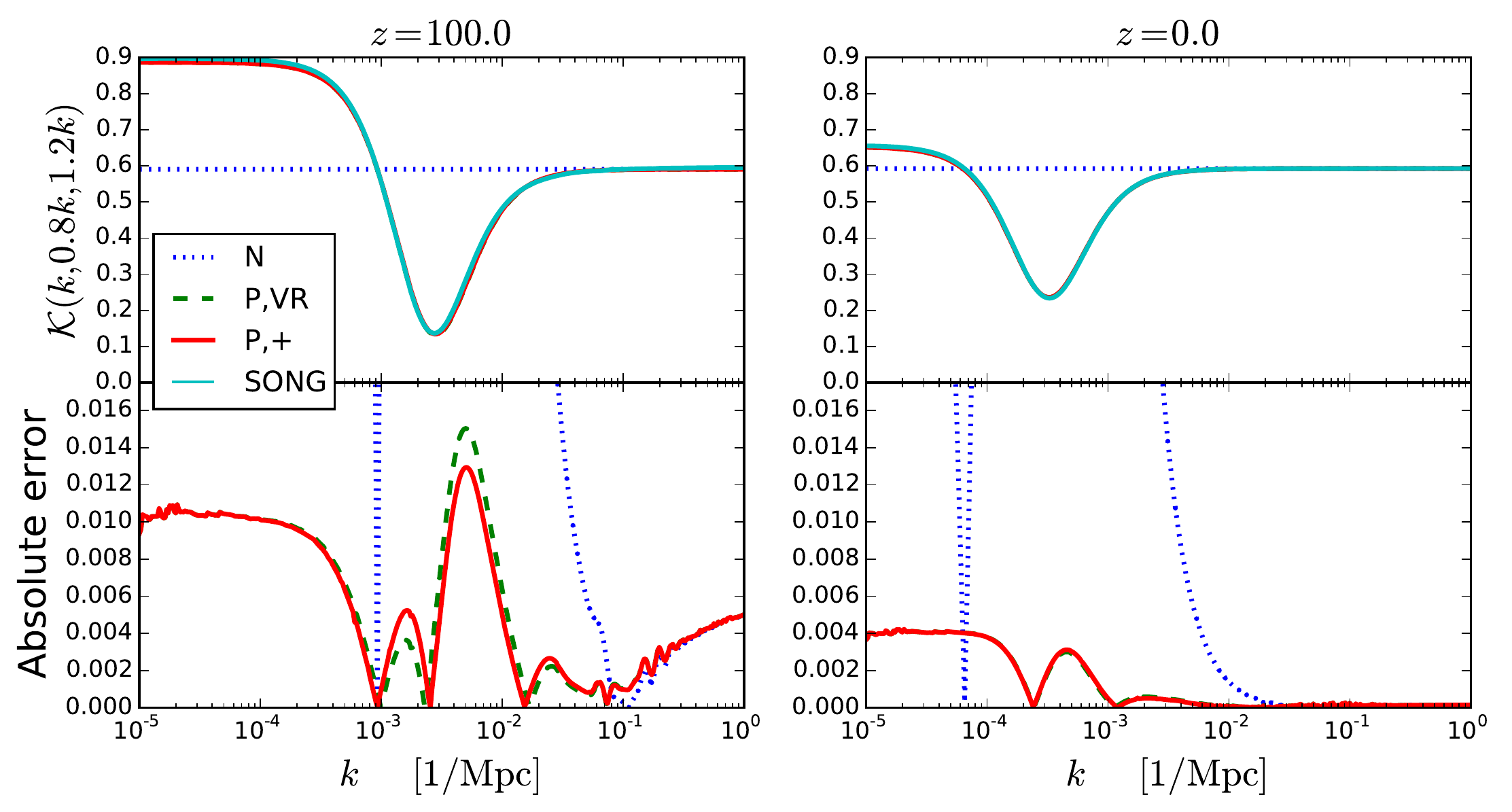}
\end{center}
\caption{The kernel computed by \SONG{} for a nearly equilateral configuration compared to the analytic estimates at redshift $z=100$ and $z=0$. 
The difference between $\mathcal{K}_{P,\text{VR}}$ and $\mathcal{K}_{P,+}$ is negligible for this configuration as we expect from the form of the correction term.}
\label{fig:simpleequi}
\end{figure} 
\begin{figure}[tb]
\begin{center}
\includegraphics[width=1.0\textwidth]{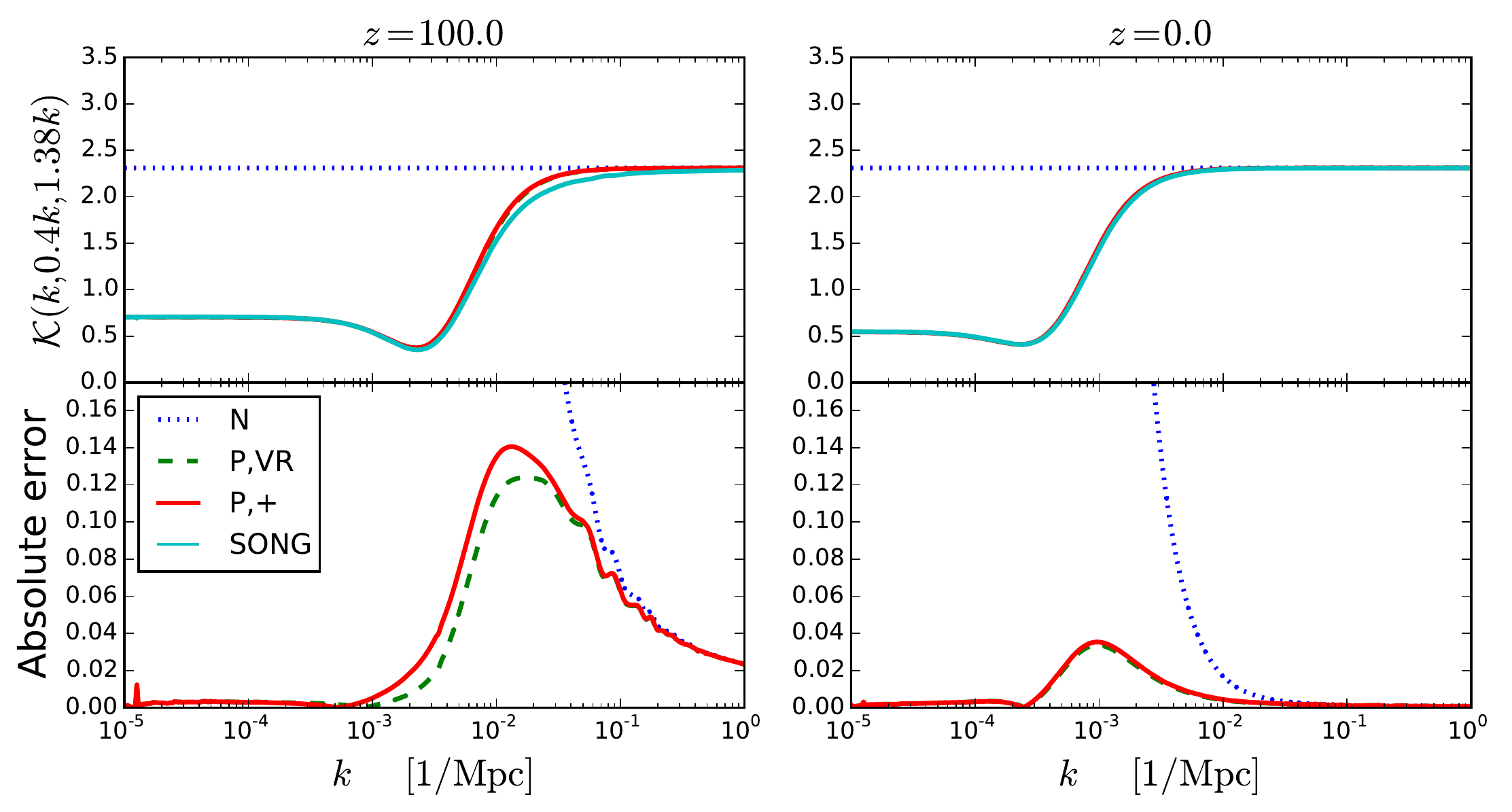}
\end{center}
\caption{The kernel computed by \SONG{} for a folded configuration compared to the analytic estimates at redshift $z=100$ and $z=0$. The GR kernels are accurate at large scales where the initial condition is correct and at small scales and late times where the Newtonian part takes over. The bump in the error at intermediate scales can be understood in this way, since the GR kernels are effectively just interpolating between these two regimes.}
\label{fig:simpleflat}
\end{figure} 
We will compare the numerically computed kernel from \SONG{} to three analytical approximations $\mathcal{K}_N$, $\mathcal{K}_{P,\text{VR}}$ and $\mathcal{K}_{P,+}$. The Newtonian kernel $\mathcal{K}_N$ is given by equation~(\ref{eq:kernelN}) and the GR kernel $\mathcal{K}_{P,\text{VR}}$ derived from~\cite{Villa:2015ppa} is given by equation~\eqref{eq:delta2p}. $\mathcal{K}_{P,+}$ is the improved kernel with the squeezed limit correction obtained by replacing 
$\gamma_{P,\text{VR}}$ by $\gamma_{P,\text{sq}}$ given by \eqref{eq:gamma_squeezed}. This is the most accurate analytic formula so we give the explicit form here:
\begin{equation} 
\mathcal{K}_{P,+} \equiv \left( \beta_{P,\text{VR}} - \alpha_{P,\text{VR}} \right) + \frac{\beta_{P,\text{VR}}}{2} \coskk \left( \frac{k_2}{k_1} + \frac{k_1}{k_2} \right) 
+ \alpha_{P,\text{VR}} \left(\coskk \right)^2 +   \gamma_{P,\text{sq}} \left( \frac{k_1}{k_2} - \frac{k_2}{k_1} \right)^2.
\label{KP+}
\end{equation}
where $\alpha_{P,\text{VR}}$ and $\beta_{P,\text{VR}}$ are given in~\eqref{eq:gammap} and $\gamma_{P,\text{sq}}$ is defined by~\eqref{eq:gamma_squeezed}. 

In figure~\ref{fig:simplesqueezed} we plot the same kernel as in figure~\ref{fig:squeezedproblem}, but now we are also comparing to $\mathcal{K}_{P,+}$. The agreement between the kernel computed by \SONG{} and $\mathcal{K}_{P,+}$ is quite remarkable, matching the baryon acoustic oscillations and improving the analytical fit by orders of magnitude. At $z=100$ the baryon perturbations and the CDM perturbations have still not equilibrated completely, which explains why the oscillations are not fit exactly\footnote{The squeezed limit computation can be generalised to compute the baryon bispectrum, employing the first-order baryon transfer functions. Instead, for simplicity, we base our computation on the Bardeen potential, using the total density contrast for baryons and dark matter.}. At $z=0$ the acoustic oscillations are well matched and the residual error is at the percent level. 

The second configuration that we have shown in figure~\ref{fig:simpleequi} is a nearly equilateral configuration. Despite not including baryons or radiation, the GR kernels are accurate at the percent level at redshift $z=100$ and at the sub-percent level today. The difference between $\mathcal{K}_{P,\text{VR}}$ and $\mathcal{K}_{P,+}$ is insignificant which confirms that the correction term in $\gamma_{P,\text{sq}}$ is safe to include in the full kernel.

Figure~\ref{fig:simpleflat} shows a folded configuration where $k_1+k_2$ is only slightly larger than $k_3$. In this case $\mathcal{K}_{P,\text{VR}}$ and $\mathcal{K}_{P,+}$ are in close agreement, again showing that the squeezed limit correction is safe to add to the full kernel. At $z=0$ the disagreement is less than a percent at large scales, but around $10^{-3}\text{Mpc}^{-1}$ we see a bump where the error goes to 4\%. At this scale 
the perturbations re-enter during radiation domination and we should consider the full intrinsic bispectrum. 
On the smaller scales the fit improves again since the solution becomes dominated by its source; the Newtonian kernel starts to become a good approximation on these scales. At $z=100$ this effect is more pronounced with the error reaching 10\% as the initial conditions still have a stronger influence. This mismatch of $\mathcal{K}_{P,\text{VR}}$ due to the intrinsic bispectrum for the folded shapes cannot be captured by a simple analytical computation, as done in $\mathcal{K}_{P,+}$ for the squeezed shapes; it requires a full second-order computation.

\begin{figure}[tb]
\begin{center}
\includegraphics[width=1.0\textwidth]{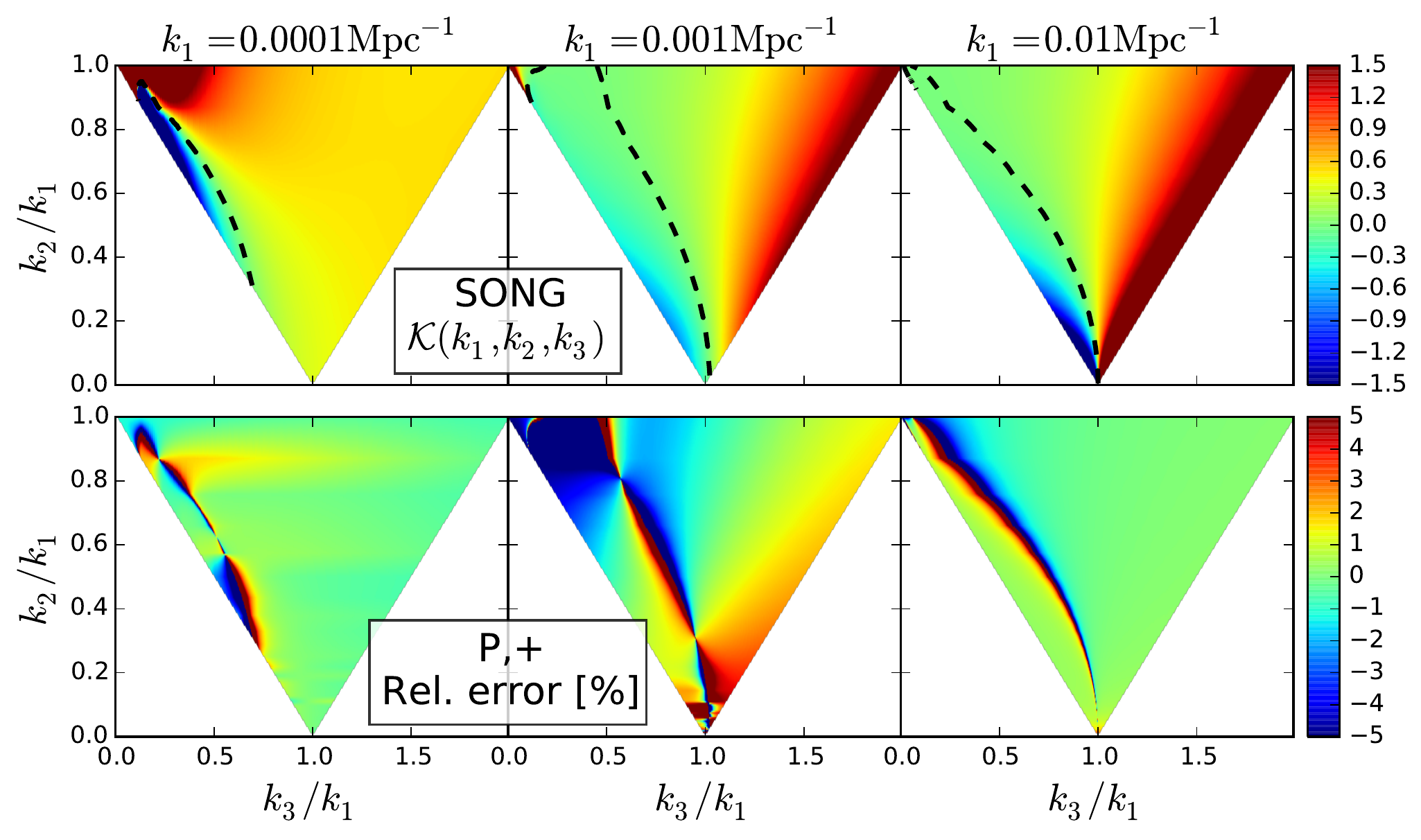}
\end{center}
\caption{The top panel shows the full matter kernel at $z=0$ computed by \SONG{} at three different scales, while the bottom panel shows the relative error of the analytic kernel $\mathcal{K}_{P,+}$ compared to the full result from \SONG{}. 
The bottom tip of the triangles corresponds to the squeezed limit ($k_2\ll k_1$), the {right edge} corresponds to folded triangles ($k_3=k_1+k_2$) and the top centre corresponds to equilateral shapes ($k_1\approx k_2\approx k_3$). 
The dashed line in the top panel shows the zero values of the kernel from \SONG{} which shows that the largest fractional error in the analytic kernel is associated with this zero-crossing.}
\label{fig:simplefullbisp}
\end{figure} 

Finally in figure~\ref{fig:simplefullbisp}, we plot the full bispectrum kernel from \SONG{} at $z=0$ together with the percentage relative error of $\mathcal{K}_{P,+}$. For three scales $k_1=\{10^{-4},10^{-3} ,10^{-2}\}\text{Mpc}^{-1}$, we plot the kernel as a function of $k_2/k_1$ and $k_3/k_1$. We have indicated the zero-crossings of $\mathcal{K}_\text{SONG}$ by a dashed line. The error of $\mathcal{K}_{P,+}$ is less than 1\% for most configurations, and the configurations where the error is bigger than 1\% are mostly associated with areas where the kernel is vanishing. The exception are the ``folded and slightly squeezed'' configurations at scale $k_1=10^{-3}\text{Mpc}$ at the lower right part of the triangle. These are configurations similar to the one shown in figure~\ref{fig:simpleflat} so at this scale we are picking up the error bump seen in that figure.

\section{Conclusion}\label{sec:conclusions} 

In this paper we have presented for the first time a full numerical calculation of the leading-order matter bispectrum obtained from the second-order Einstein-Boltzmann code \SONG{} originally developed to study nonlinear evolution in the CMB \cite{Pettinari:2013he,pettinari:2015a,Pettinari:2014iha}. 
The full bispectrum includes the usual nonlinear Newtonian evolution during the matter era \cite{Bernardeau:2001qr,Noh:2004bc}, but also the general relativistic initial conditions due to the nonlinear constraint equations of general relativity \cite{Bartolo:2005xa,Bartolo:2010rw,Bruni:2014xma}. It includes the effects of radiation coupled to baryons at early times \cite{Fitzpatrick:2009ci} and the cosmological constant at late times \cite{Bruni:2013qta,Uggla:2013kya,Villa:2015ppa}.

Measurements of the large-scale galaxy 3-point function already indicate the presence of baryon acoustic oscillation \cite{Gaztanaga:2008sq,Slepian:2015hca}. 
We have shown that baryon acoustic oscillations come into the matter bispectrum not just through the variance of the modes themselves (e.g. \cite{Sefusatti:2006pa}), but also through how the modes are correlated.  That is, while BAO features naturally arise in the bispectrum from the power spectrum factors in equation (\ref{redbispectrum}), we have shown for the first time that they also arise in the kernel itself through the second-order perturbation.  These features are particularly strong in squeezed bispectrum configurations, as is shown in figure~\ref{fig:squeezedproblem}.

We have also derived a novel analytical approximation for the total matter bispectrum~(\ref{KP+}), valid in both the matter- and $\Lambda$-dominated eras. It accurately reproduces relativistic effects in the Poisson gauge~\cite{Villa:2015ppa} and, in the squeezed limit, the feedback of photons and baryons on CDM induced by the radiation-dominated era, faithfully reproducing the baryon acoustic oscillations at low redshift, see figure~\ref{fig:simpleequi}. 
As shown in figure~\ref{fig:simplefullbisp}, our analytic approximation is reliable at the percent level for most configurations at the present day.

This work is an important step towards making consistent relativistic predictions for upcoming large-scale structure surveys such as Euclid and SKA. 
A numerical Boltzmann code such as \SONG{} allows us to compute separately the cold dark matter and baryon density contrast up to second order and hence is a necessary step towards constructing consistent relativistic initial conditions for $N$-body codes at high redshift and/or on large scales.

We have recently shown how to set up relativistic initial conditions for N-body simulations at first order in general relativity that can be consistently evolved using standard Newtonian equations of motion~\cite{Fidler:2015npa} (see also~\cite{Chisari:2011iq,Rigopoulos:2013nda}). However no such consistent N-body treatment yet exists at second order in general relativity. An attempt has been made to incorporate post-Newtonian corrections in N-body evolution \cite{Adamek:2015eda}, but this approach remains linear on large scales (however see~\cite{Milillo:2015cva}). The development of a second order relativistic treatment for N-body simulations remains an outstanding problem before the results of this work can be used to obtain reliable predictions for future galaxy surveys.

\acknowledgments
We wish to thank Marco Bruni, Cornelius Rampf, Eleonora Villa and Matteo Tellarini for several enlightening discussions. TT, RC, KK and DW are supported by the UK Science and Technology Facilities Council grants ST/K00090X/1. KK is supported by the European Research Council grant through 646702 (CosTesGrav). CF is supported by the Wallonia-Brussels Federation grant ARC11/15-040 and the Belgian Federal Office for Science, Technical \& Cultural Affairs through the Interuniversity Attraction Pole P7/37.

\appendix

\section{Newtonian Boltzmann equation in comoving coordinates}\label{sec:NewtonianBoltzmann}
The total differential of the dark matter distribution function is
\begin{align}
\di f(\tau, \vec{x},\vec{u}) &= \diffpar{f}{\tau} \di \tau + \diffpar{f}{\vec{x}} \cdot \di \vec{x} +  \diffpar{f}{\vec{u}} \cdot \di \vec{u} \\
&= \diffpar{f}{\tau} \di \tau + \diffpar{f}{\vec{x}} \cdot \diff{\vec{x}}{\tau} \di \tau + \diffpar{f}{\vec{u}} \cdot \diff{\vec{u}}{\tau} \di \tau \\
&= \left[\diffpar{f}{\tau}  + \diffpar{f}{\vec{x}} \cdot \vec{u} + \diffpar{f}{\vec{u}} \cdot \left( \diff{\vec{v}}{\tau} - \Hcdot \vec{x} \right) \right] \di \tau.
\end{align}
In Newtonian theory, the equation of motion of a point particle in a gravitational field $\Phi_N$ is
\begin{equation}
\left. \diff{\vec{v}}{t} \right|_\text{matter} = - \diff{\Phi_N}{\vec{r}}.
\end{equation}
However, in GR the presence of a cosmological constant provides an isotropic acceleration independent of matter. Taking the time derivative of equation ~\eqref{eq:vpeculiar} with $\Omega_m=0$ yields
\begin{equation}
\left. \diff{\vec{v}}{\tau} \right|_\Lambda = H_0^2 a^2 \Omega_\Lambda \vec{x}.
\end{equation}
The assumption that the particles only interact through gravity, $\diff{f}{\tau}=0$, thus leads to the collision-less Boltzmann equation
\begin{align}
\diffpar{f}{\tau}  &= - \diffpar{f}{\vec{x}} \cdot \vec{u} - \diffpar{f}{\vec{u}} \cdot \left( - \diff{\Phi_N}{\vec{x}} +H_0^2 a^2 \Omega_\Lambda \vec{x} - \Hcdot \vec{x} \right) \\
&= - \diffpar{f}{\vec{x}} \cdot \vec{u} + \diffpar{f}{\vec{u}} \cdot \diff{\Phi_E}{\vec{x}}, \label{eq:Boltzmann}
\end{align}
where we have introduced an effective potential
\begin{equation}
\Phi_E \equiv \Phi_N + \frac{1}{2} \left( \Hcdot - H_0^2 \Omega_\Lambda  a^2 \right) x^2.\label{eq:effective}
\end{equation}
We can write the dark matter density in terms of the density contrast and the mean density,
\begin{equation}
\rho_m(\tau,\vec{x}) \equiv \left(1+\delta(\tau,\vec{x}) \right) \bar{\rho}_m.
\end{equation}
By taking the Laplacian (in $\vec{x}$-space) of equation~\eqref{eq:effective} we can derive a Laplace equation for $\Phi_E$:
\begin{align}
\nabla^2 \Phi_E &= \nabla^2 \Phi_N + 3   \left( \Hcdot - H_0^2 \Omega_\Lambda a^2 \right) \\
&= 4\pi G a^2 \left(1+\delta(\tau,\vec{x}) \right) \bar{\rho}_m + 3 H_0^2 \left( \Omega_m a^{-1} + \Omega_\Lambda a^2 - \frac{3}{2} \Omega_m a^{-1} -  \Omega_\Lambda  a^2 \right) \\
&= \frac{3}{2} \HcO a^{-1} \delta(\tau,\vec{x}).
\end{align}
This shows that $\Phi_E$ vanishes when the matter distribution is exactly homogeneous. 

To relate the distribution function to physical fluid variables, we will need the coordinate transformation back to physical phase space. From equation~\eqref{eq:vpeculiar} we have
\begin{equation}
\begin{pmatrix}
\vec{r} \\
\vec{p}
\end{pmatrix}
=
\begin{pmatrix}
a & 0\\
m\Hc & m
\end{pmatrix}
\begin{pmatrix}
\vec{x} \\
\vec{u}
\end{pmatrix},
\end{equation}
where the $6\times6$ matrix has been written in block form. The differentials transform as the Jacobian of this transformation, so we immediately find
\begin{equation}
\di N =  f(\vec{r},\vec{p})  \di ^3 \vec{r} \di ^3 \vec{p}= \begin{vmatrix} a & 0\\
m\Hc & m
\end{vmatrix}
 f(\vec{x},\vec{u})  \di ^3 \vec{x} \di ^3 \vec{u} = a^3 m^3  f(\vec{x},\vec{u})   \di ^3 \vec{x} \di ^3 \vec{u}.
\end{equation}
The number density in the comoving coordinates is now
\begin{equation}
n(\tau,\vec{x}) =  \int \di^3 \vec{u} \frac{\di N}{\di^3 \vec{x}} = \int \di^3 \vec{u}  a^3 m^3  f(\vec{x},\vec{u}) \equiv \kinavg{f},
\end{equation}
where we introduced a short hand notation for the velocity integral. The next two velocity moments define the peculiar velocity flow $\velo$ and the stress tensor $\sigma_{ij}$ respectively:
\begin{align}
 \velo(\tau,\vec{x}) &= \frac{1}{n(\tau,\vec{x}) }  \kinavg{ \vec{u} f }, \\
\sigma_{ij}(\tau,\vec{x}) &= \frac{1}{n} \kinavg{ u_i u_j f } -  \velo_i \velo_j.
\end{align}
We will now derive the fluid equations by taking kinetic moments of the Boltzmann equation~\eqref{eq:Boltzmann}. Since both the phase space volume $\di ^3 \vec{r}  \di ^3 \vec{p}$ and the comoving volume $\di^3 \vec{x}$ is conserved, the quantity $a^3 m^3 \di ^3 \vec{u}$ is conserved as well and $\vec{u}\propto \frac{1}{a}$. This also implies that our definition of the kinetic averaging is time independent,
\begin{equation}
\kinavg{\diffpar{g}{\tau}} = \diffpar{}{\tau} \kinavg{g}.
\end{equation}
The first moment gives
\begin{align}
\kinavg{\diffpar{f}{\tau}}  &= - \kinavg{\diffpar{f}{\vec{x}} \cdot \vec{u}} + \kinavg{\diffpar{f}{\vec{u}} \cdot \diff{\Phi_E}{\vec{x}}} \\
\diffpar{}{\tau} \kinavg{f} &= -\diffpar{}{x^j} \kinavg{ u^j f} + \diff{\Phi_E}{x^j} \kinavg{\diffpar{f}{u^j}} \\
\dot{n} &= -\diffpar{}{x^j} \left( n \velo^j \right) \\
\dot{\delta} &= -\diffpar{}{x^j} \left[ (1+\delta) \velo^j \right], \label{eq:ContReal}
\end{align}
where we used $n = \rho/m = (1+\delta) \bar{\rho}/m$. This is the continuity equation which just states that the number of particles is conserved. The second moment gives
\begin{align}
\kinavg{u^i \diffpar{f}{\tau}}  &= - \kinavg{u^i \diffpar{f}{\vec{x}} \cdot \vec{u}} + \kinavg{u^i \diffpar{f}{\vec{u}} \cdot \diff{\Phi_E}{\vec{x}}} \nonumber \\
\kinavg{\diffpar{u^i f}{\tau}- \diffpar{u^i}{\tau}f} &= -\diffpar{}{x^j} \kinavg{u^i u^j f} + \diff{\Phi_E}{x^j} \kinavg{u^i \diffpar{f}{u_j}} \nonumber \\
\diffpar{}{\tau} \kinavg{u^i f} + \Hc \kinavg{u^i f} &=  -\diffpar{}{x^j} \left[ \sigma^{ij}(\tau,\vec{x}) + n(\tau,\vec{x}) \velo^i \velo^j \right] - \diff{\Phi_E}{x^j} \kinavg{\delta^{ij} f} \nonumber \\
n\dot{\velo^i} -  \diffpar{}{x^j} \left( n \velo^j \right) \velo^i  + \Hc n \velo^i &= -\diffpar{}{x^j} \left[ n \sigma^{ij}(\tau,\vec{x}) + n(\tau,\vec{x}) \velo^i \velo^j \right] - \diff{\Phi_E}{x^i} n \nonumber \\
\dot{\velo^i}  &= - \Hc \velo^i -\velo^j \diffpar{\velo^i}{x^j} - \frac{1}{n}\diffpar{}{x^j} n \sigma^{ij}  - \diff{\Phi_E}{x^i}. \label{eq:EulerReal}
\end{align}
From now on we will neglect the stress tensor $\sigma_{ij}$, and we will assume that the vorticity of $\velo$ vanish, i.e. $\velo$ is fully described by a scalar  potential. The latter assumption is consistent with neglecting the stress tensor since one can show that vorticity is sourced by $\sigma_{ij}$. In terms of the velocity potential $\velpot$ defined as $\velo = \nabla \velpot$, equation~\eqref{eq:ContReal} and~\eqref{eq:EulerReal} read
\begin{align}
\dot{\delta} &= -\partial_j \left[ (1+\delta) \partial_j \velpot \right] \\
\partial_i \dot{\velpot} &= - \Hc \partial_i \velpot - \partial_j \velpot \partial_j \partial_i \velpot - \partial_i \Phi_E \\
\dot{\theta} &= - \Hc \theta - \partial_i \partial_j \nabla^{-2}\theta \partial_j \partial_i \nabla^{-2}\theta - \partial_j \nabla^{-2} \theta \partial_j \theta - \nabla^2 \Phi_E.
\end{align}

\section{$\Lambda$CDM growth functions}\label{sec:growth}
We will now briefly relate some basic results concerning the first and second-order growth functions $D(a)$ and $F(a)$ as well as $\Hc E(a)$. $D$ and $F$ are the fastest growing solutions of equation~\eqref{eq:Dequation} and equation~\eqref{eq:Fequation} respectively, and $\Hc E(a)$ arises in the coordinate transformation in $\Lambda$CDM that absorbs the long mode and it is defined in equation~\eqref{eq:Eequation}.  It will be convenient for us to define 
\begin{equation}
x\equiv \frac{\Omega_\Lambda}{\Omega_m} a^3,
\end{equation}
and 
we use the fact that any hypergeometric function that admits a quadratic transformation can be expressed in terms of the Legendre function.

\begin{align}
D(a) &= \frac{5}{2} \HcO \frac{\Hc}{a} \int_0^a  \frac{da'}{\Hc^3(a')}, \label{eq:Dintegral} \\
&= a\sqrt{1+x} \hypF{\frac{3}{2}}{\frac{5}{6}}{\frac{11}{6}}{-x}, \nonumber \\
&= a \hypF{\frac{1}{3}}{1}{\frac{11}{6}}{-x}, \nonumber \\
&= 2^\frac{5}{6} \Gamma\left(\frac{11}{6}\right) a x^{-\frac{5}{12}} P_{-\frac{7}{6}}^{-\frac{5}{6}}\left(\sqrt{1+x}\right). \nonumber \\
\Hc E(a) &= \Hc \sqrt{\frac{a}{\Omega_m}} \frac{1}{a^2} \int_0^a \frac{a'}{\Hc(a')} da',\\
&= -\frac{2}{5} \sqrt{1+x} \hypF{\frac{1}{2}}{\frac{5}{6}}{\frac{11}{6}}{-x}, \label{eq:Ehyper} \nonumber \\ 
&= -\frac{2}{5}  2^\frac{5}{6}  \Gamma\left(\frac{11}{6}\right) \sqrt{1+x}  x^{-\frac{5}{12}} P_{-\frac{5}{6}}^{-\frac{5}{6}}\left(\sqrt{1+x}\right).  \nonumber \\
\Hc \eta(a) &= \frac{\Hc}{H_0} \int_0^a \frac{1}{a' \Hc(a')} da', \\ 
&=\sqrt{1+x} \hypF{\frac{1}{6}}{\frac{1}{2}}{\frac{7}{6}}{-x}. \nonumber 
\end{align}

\begin{figure}\label{fig:fuvw}
\begin{center}
\includegraphics[width=1.0\textwidth]{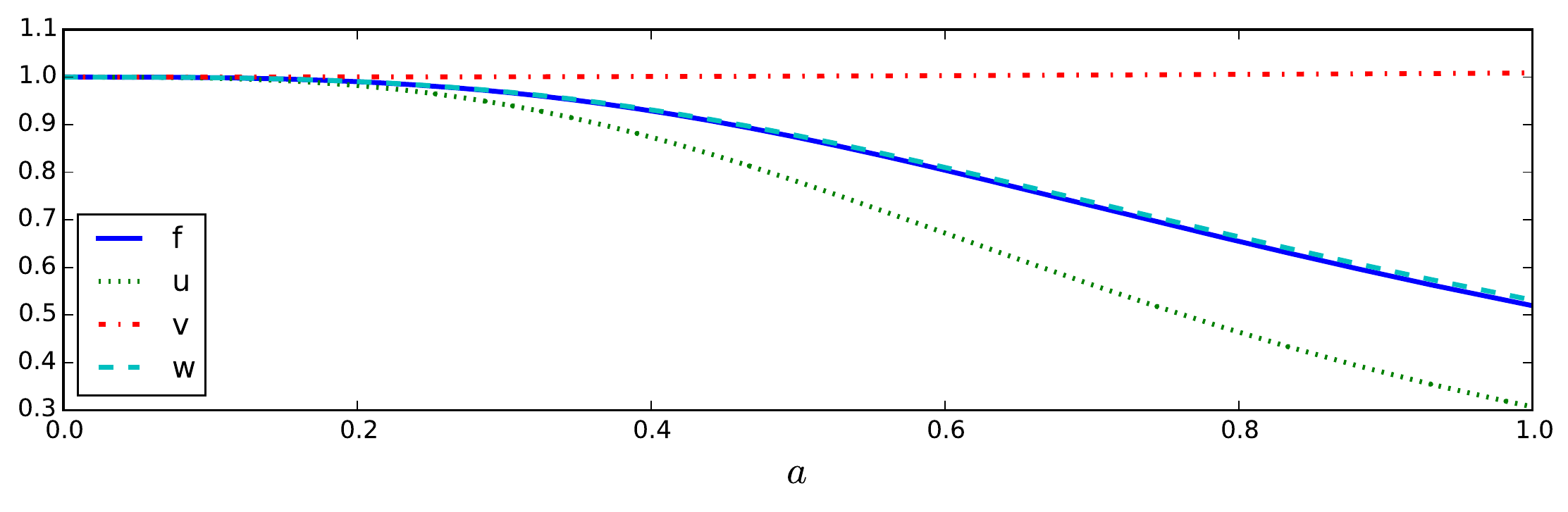}
\end{center}
\caption{Redshift dependence of the four functions $f$, $u$, $v$ and $w$ that fully describe the deviations from EdS of the second-order density kernel. Note that $v\simeq 1$ and that $w\simeq f$.}
\end{figure}

We are not aware of any explicit expression for $F$ in $\Lambda$CDM, but note that $F\xrightarrow{EdS}\frac{3}{7}a^2$. In order to make the EdS limit obvious, we define 4 functions that all evaluate to $1$ in EdS:
\begin{align}
f &= \frac{\dot{D}}{D\Hc} = \frac{a D'(a)}{D}, & u &\equiv \frac{2}{3} \left[1-\frac{\Hcdot}{\Hc^2}\right] = \frac{1}{1+\frac{\Omega_\Lambda}{\Omega_M} a^3}, \\
v &\equiv \frac{7F}{3D^2}, & w &\equiv \frac{7\dot{F}}{6\Hc D^2} =  \frac{7aF'(a)}{6D^2}.
\end{align}
We have shown the redshift evolution of these quantities in figure~\ref{fig:fuvw}.

\section{Separate universe approach} \label{sec:separate}
In this appendix we derive the local coordinate transformation (\ref{eq:Eequation}) to remove the long-wavelength curvature perturbation $\zeta$, as well as the relation
\begin{equation}
\Hc E = -\frac{1}{1+\frac{3u}{2f}}.
\end{equation} 
We perform the following {\it local} coordinate transformation:
\begin{equation}
\tilde{\eta} = \eta + \epsilon(\eta), \quad 
\tilde{x}^i = x^i (1 + \lambda). 
\end{equation}
Note that the rescaling of spatial coordinates is not the usual gauge transformation and this transformation leaves the line element equation~(\ref{firstPoisson}) in the same form only locally where we can ignore the spatial dependence of $\lambda$. As we are interested in a patch whose size is smaller than the long-wavelenght mode, we can perform this rescaling if $\lambda$ is associated with the long-wavelength curvature perturbations $\zeta_l$.   
The metric perturbations $\Psi$ and $\Phi$ are transformed as 
\begin{equation}
\tilde{\Phi} = \Phi - \Hc \epsilon - \lambda, \quad 
\tilde{\Psi} = \Psi - \Hc \epsilon -\dot{\epsilon}. 
\label{eq:trans}
\end{equation}
The comoving curvature perturbation is given by equation~(\ref{zeta})
\begin{equation}
\zeta  = \Phi - \frac{2}{3} \frac{1}{1+w} \left[ \Psi - \frac{a}{\dot{a}} \dot{\Phi} \right].
\end{equation}
Using equation~(\ref{eq:trans}), we can show that the comoving curvature perturbation transforms as 
\begin{equation}
\tilde{\zeta} = \zeta - \lambda. 
\end{equation}
Thus by choosing $\lambda = \zeta$, we can remove the long-wavelength comoving curvature perturbation in the local patch. In the absence of anisotropic stress, the metric perturbations satisfy $\tilde{\Phi} + \tilde{\Psi} = \Phi + \Psi =0$. This gives a condition on $\epsilon$ as 
\begin{equation}
\dot{\epsilon} + 2 \Hc \epsilon = - \zeta.
\end{equation}
The solution to this equation gives equation~(\ref{eq:Eequation}). 

Since the comoving curvature perturbation vanishes, $\tilde{\Phi}$ also vanishes in the local patch. This gives the following equation:
\begin{equation}
\tilde{\Phi} = \Phi- \Hc \epsilon + \zeta =0. 
\end{equation}
Defining $\epsilon = E(\eta) \zeta$, we obtain 
\begin{equation}
\Hc E = - \frac{1}{1+ \frac{3 u}{2 f}},
\end{equation}
where we used the fact that the comoving curvature perturbation can be rewritten as 
\begin{equation}
\zeta = \left(1 + \frac{2 f}{3 u} \right) \Phi. 
\label{eq:zetaC}
\end{equation}

\section{Fourier dictionary}\label{sec:dict}
We will use GR results presented in Ref.~\cite{Villa:2015ppa}. They write their formulae in terms of the present day linearly extrapolated Newtonian potential, $\phi_0\equiv -({a_0}/{D_0})\Phi_0$, where $a_0$ is the present day scale factor and $D_0$ the present growth factor. This section contains a convenient dictionary for converting second-order real-space results into Fourier space. We will use $\FourierMap$ to denote a mapping under the Fourier operator defined in equation~\eqref{eq:FourierDef}.

\begin{align}
\phi_0^2 &\FourierMap \KernelPhi{} \\
\left( \nabla^2 \phi_0\right) ^2 &\FourierMap \KernelPhi{k_1^2k_2^2} \\
\left( \nabla \phi_0\right) ^2 &\FourierMap \KernelPhi{(-1)k_1 k_2 \coskk} \\
2\phi_0^{,l}\nabla^2 \phi_{0,l} &\FourierMap \KernelPhi{k_1^2k_2^2 \coskk \left( \frac{k_2}{k_1} + \frac{k_1}{k_2} \right)} \\
2\phi_0\nabla^2 \phi_0 &\FourierMap \KernelPhi{(-1) \left( k_1^2 + k_2^2\right) } \\
\phi_{0,lm} \phi_0^{,lm} &\FourierMap \KernelPhi{k_1^2k_2^2 (\coskk)^2 } \\
\nabla^{-2}(\phi_0^{,l}\phi_{0,l}) &\FourierMap \KernelPhi{\frac{k_1k_2}{k^2} \coskk} \\
\nabla^{-4} \left( \phi_0^{,l}\phi_0^{,m} \right)_{,lm} &\FourierMap \KernelPhi{\frac{k_1^2k_2^2}{k^4} \left[ 1 + \left( \frac{k_1}{k_2} + \frac{k_2}{k_1} \right) \coskk + \left(\coskk\right)^2 \right]  }\\
\nabla^{-2}\left[ \left( \nabla^2  \phi_0 \right)^2 \right] &\FourierMap \KernelPhi{\frac{-k_1^2k_2^2}{k^2}}\\
\nabla^{-2}\left[ \phi_{0,ik} \phi_0^{,ik} \right] &\FourierMap \KernelPhi{\frac{-k_1^2k_2^2}{k^2} \left(\coskk\right)^2 }
\end{align}
From this dictionary we also find:
\begin{align}
\Theta_0 &\equiv \frac{1}{2} \nabla^{-2} \left[ \frac{1}{3} \phi_0^{,l} \phi_0^{,l} - \nabla^{-2} \left( \phi_0^{,l} \phi_0^{,m} \right)_{,lm} \right]  \\
&\FourierMap \KernelPhi{\left( \frac{1}{6} \frac{k_1k_2}{k^2}  \coskk - \frac{1}{2} \frac{k_1^2k_2^2}{k^4} \left[ 1 + \left( \frac{k_1}{k_2} + \frac{k_2}{k_1} \right) \coskk + \left(\coskk\right)^2 \right]  \right)} \nonumber \\
\Psi_0 &\equiv -\frac{1}{2} \nabla^{-2} \left[ \left(\nabla^2 \phi_0 \right)^2 - \phi_{0,ik} \phi_0^{,ik} \right]  \\
&\FourierMap \KernelPhi{ \frac{1}{2} \frac{k_1^2k_2^2}{k^2} \left[1-\left(\coskk\right)^2 \right]  } \nonumber
\end{align}
It will be useful to write these kernels in terms of $\delta_P^{(1)}$. We will pull out factors of $k$ to make the remaining part dimensionless w.r.t. $k_i$ and we will write any remaining $k$'s in terms of $k_1$, $k_2$ and $\coskk$ using the relations
\begin{align}
\frac{k^2}{k_1k_2} &= \frac{k_1}{k_2} + \frac{k_2}{k_1} + 2 \coskk, \\
\frac{k^4}{k_1^2k_2^2} &= 4 \left[1+ \coskk\left( \frac{k_1}{k_2}+\frac{k_2}{k_1}\right) +\left(\coskk\right)^2\right] + \left( \frac{k_1}{k_2}-\frac{k_2}{k_1}\right)^2.
\end{align}
\paragraph{Terms proportional to $\frac{1}{k^4}$:}
\begin{align}
\phi_0^2 &\FourierMap \KernelDelta{D^2k^4}{\frac{k^4}{k_1^2k_2^2}} \\
&=  \KernelDeltaBis{D^2k^4}{ \left( \left[ \frac{k_1}{k_2}-\frac{k_2}{k_1}\right]^2 \right. } + \nonumber \\
&\qquad \left. \left. + 4 \left[1+ \coskk\left( \frac{k_1}{k_2}+\frac{k_2}{k_1}\right) +\left(\coskk\right)^2\right] \right) \right\}  \nonumber \\
\Theta_0 &\FourierMap \KernelDeltaBis{D^2k^4}{\left( \frac{k^2}{6k_1k_2}  \coskk \right. } -  \\
&\qquad \left. \left. - \frac{1}{2} \left[ 1 + \left( \frac{k_1}{k_2} + \frac{k_2}{k_1} \right) \coskk + \left(\coskk\right)^2 \right]  \right) \right\} \nonumber \\
&=  \KernelDeltaBis{D^2k^4}{ \left( - \frac{1}{2} - \frac{1}{3} \left( \frac{k_1}{k_2} + \frac{k_2}{k_1} \right) \coskk \right. } - \nonumber \\
&\qquad \left. \left.  - \frac{1}{6}\left(\coskk\right)^2  \right) \right\}  \nonumber
\end{align}
\paragraph{Terms proportional to $\frac{1}{k^2}$:}
\begin{align}
\left( \nabla \phi_0\right) ^2 &\FourierMap \KernelDelta{D^2k^2}{(-1)\frac{k^2}{k_1k_2} \coskk}   \\
&=  \KernelDelta{D^2k^2}{(-1) \left( \left[\frac{k_1}{k_2} + \frac{k_2}{k_1} \right] \coskk + 2 \left(\coskk\right)^2 \right)} \nonumber \\
\Psi_0 &\FourierMap \KernelDelta{D^2k^2}{ \frac{1}{2} \left[1-\left(\coskk\right)^2 \right]  }\\
2\phi_0\nabla^2 \phi_0 &\FourierMap \KernelDelta{D^2k^2}{(-1)\frac{k^2}{k_1}{k_2} \left(\frac{k_1}{k_2}+\frac{k_2}{k_1}\right)}  \\
&= \KernelDelta{D^2k^2}{(-1)\left( \left[ \frac{k_1}{k_2} - \frac{k_2}{k_1} \right]^2 + 4 + 2\coskk\left[ \frac{k_1}{k_2} + \frac{k_2}{k_1} \right] \right) } \nonumber
\end{align}
\paragraph{Terms proportional to $\frac{1}{k^0}$:}
\begin{align}
\left( \nabla^2 \phi_0\right) ^2 &\FourierMap \KernelDelta{D^2}{} \\
2\phi_0^{,l}\nabla^2 \phi_{0,l} &\FourierMap \KernelDelta{D^2}{\coskk \left( \frac{k_2}{k_1} + \frac{k_1}{k_2} \right)} \\
\phi_{0,lm} \phi_0^{,lm} &\FourierMap \KernelDelta{D^2}{ (\coskk)^2 } 
\end{align}

\bibliographystyle{JHEP}
\bibliography{references}

\end{document}